%% file: TCOM-TPS-21-1336.R2_RIS-RQSSK_MDinan_NPerovic_MFlanagan.tex
\documentclass[twocolumn,journal]{IEEEtran}
\usepackage[T1]{fontenc}
\usepackage[latin9]{inputenc}
\usepackage{units}
\usepackage{enumitem}
\usepackage{amsmath}
\usepackage{amsthm}
\usepackage{amssymb}
\usepackage{graphicx}
\usepackage[unicode=true,
 bookmarks=true,bookmarksnumbered=true,bookmarksopen=true,bookmarksopenlevel=1,
 breaklinks=false,pdfborder={0 0 0},pdfborderstyle={},backref=false,colorlinks=false]
 {hyperref}
\hypersetup{pdftitle={Your Title},
 pdfauthor={Your Name},
 pdfpagelayout=OneColumn, pdfnewwindow=true, pdfstartview=XYZ, plainpages=false}

\makeatletter
\newlist{casenv}{enumerate}{4}
\setlist[casenv]{leftmargin=*,align=left,widest={iiii}}
\setlist[casenv,1]{label={{\itshape\ \casename} \arabic*.},ref=\arabic*}
\setlist[casenv,2]{label={{\itshape\ \casename} \roman*.},ref=\roman*}
\setlist[casenv,3]{label={{\itshape\ \casename\ \alph*.}},ref=\alph*}
\setlist[casenv,4]{label={{\itshape\ \casename} \arabic*.},ref=\arabic*}
\theoremstyle{plain}
\newtheorem{thm}{\protect\theoremname}

\usepackage[caption=false,font=footnotesize]{subfig}
\usepackage{acronym}
\AtBeginDocument{\input{acro.tex}}
\usepackage{cite}

\makeatother

\providecommand{\casename}{Case}
\providecommand{\theoremname}{Theorem}

\begin{document}
\title{RIS-Assisted Receive Quadrature Space-Shift Keying: A New Paradigm
and Performance Analysis}
\author{Mohamad H. Dinan,~\IEEEmembership{Member,~IEEE,} Nemanja Stefan
Perovi\'c,~\IEEEmembership{Member,~IEEE,}\\and~Mark F. Flanagan,~\IEEEmembership{Senior~Member,~IEEE}\thanks{This work was funded by the Irish Research Council (IRC) under the
Consolidator Laureate Award Programme (grant number IRCLA/2017/209).}\thanks{Mohamad H. Dinan and Mark F. Flanagan are with the School of Electrical
and Electronic Engineering, University College Dublin, Belfield, Dublin
4, D04 V1W8 Ireland (email: \protect\href{mailto:mohamad.hejazidinan@ucdconnect.ie}{mohamad.hejazidinan@ucdconnect.ie};
\protect\href{mailto:mark.flanagan@ieee.org}{mark.flanagan@ieee.org}).}\thanks{Nemanja Stefan Perovi\'c was with the School of Electrical and Electronic
Engineering, University College Dublin, Belfield, Dublin 4, D04 V1W8
Ireland. He is now with Universit\'e Paris-Saclay, CNRS, CentraleSup\'elec,
Laboratoire des Signaux et Syst\`emes, 3 Rue Joliot-Curie, 91192
Gif-sur-Yvette, France (email: \protect\href{mailto:nemanja-stefan.perovic@centralesupelec.fr}{nemanja-stefan.perovic@centralesupelec.fr}).}}
\maketitle
\begin{abstract}
\Acp{RIS} represent a promising candidate for \ac{6G} wireless networks,
as the \ac{RIS} technology provides a new solution to control the
propagation channel in order to improve the efficiency of a wireless
link through enhancing the received signal power. In this paper, we
propose \ac{RIS-RQSSK}, which enhances the spectral efficiency of
an \ac{RIS}-based \ac{IM} system by using the real and imaginary
dimensions independently for the purpose of \ac{IM}. Therefore, the
error rate performance of the system is improved as \emph{all} \ac{RIS}
elements reflect the incident transmit signal toward \emph{both}
selected receive antennas. At the receiver, a low-complexity but effective
\ac{GD} can be employed which determines the maximum energy per dimension
at the receive antennas. A max-min optimization problem is defined
to maximize the received \ac{SNR} components at both selected receive
antennas; an analytical solution is provided based on Lagrange duality.
In particular, the multi-variable optimization problem is shown to
reduce to the solution of a single-variable equation, which results
in a very simple design procedure. In addition, we investigate the
\ac{ABEP} of the proposed \ac{RIS-RQSSK} system and derive a closed-form
approximate upper bound on the \ac{ABEP}. We also provide extensive
numerical simulations to validate our derivations. Numerical results
show that the proposed \ac{RIS-RQSSK} scheme substantially outperforms
recent prominent benchmark schemes. This enhancement considerably
increases with an increasing number of receive antennas.\acresetall{}
\end{abstract}


\section{Introduction\label{sec:Introduction}}

In wireless communications, the recent emergence of many new applications
and services necessitates the advent of new technologies in order
to support a very large number of mobile devices as well as massive
machine-type communications. To this end, several new technologies
have emerged in \ac{5G} wireless networks, including \ac{mMIMO},
\ac{mmWave} communications and small cells. However, these technologies
intensify the use of energy and the hardware cost, which makes practical
targets difficult to reach. In addition, \ac{5G} appears to be insufficient
to meet the forthcoming requirements of next-generation wireless communications,
as \ac{5G} technologies only target the endpoints of a wireless link,
and make no attempt to influence or design the wireless \emph{environment}
which plays a major role in degrading the link's efficiency. Therefore,
controlling the propagation channel has received growing attention
in the last few years, and accordingly the technology of \acp{RIS}
is a potentially key approach for \ac{6G} wireless networks \cite{basar2019wireless,di2019smart,gacanin2020wireless,di2020smart,alghamdi2020intelligent,liu2021reconfigurable,liang2021reconfigurable,wu2021intelligent}.

\Acp{RIS} are electromagnetic surfaces that can be electronically
controlled by the network operator. An \ac{RIS} consists of an array
of small, low-cost and nearly-passive scattering elements that can
induce a pre-designed phase shift in the incident wave. Thus \acp{RIS}
can modify in an energy-efficient manner the scattering, reflection
and refraction of the environment with a view to enhancing the efficiency
of a wireless network.

In \cite{basar2019transmission}, an \ac{RIS} was deployed to enhance
a communication system in two different scenarios: an \ac{RIS}-based
\ac{SISO} scheme, where an \ac{RIS} acts as a reflector and helps
to perform beamforming, and an \ac{RIS-AP} scheme, in which an \ac{RIS}
is utilized as a transmitter (access point) to convey the information
bits through the phases of the \ac{RIS} elements. For each scheme,
the \ac{SEP} of the system was analyzed and an enormous benefit was
demonstrated compared to the conventional wireless system without
the \ac{RIS}. An \ac{RIS} was deployed in a \ac{MISO} system in
\cite{lin2020reconfigurable}, where the \ac{RIS}, in addition to
beamforming, also conveys its own information data via \ac{RPM}.
An \ac{AO} algorithm was used to optimize active and passive beamforming,
respectively, at the transmitter and \ac{RIS}, in order to maximize
the signal power at the receiver. A multi-user \ac{RIS}-based downlink
communication system was considered in \cite{huang2019reconfigurable}
and an optimization problem was defined to maximize the energy efficiency
of the system; again, an \ac{AO} algorithm was applied to optimize
both the transmit power allocation and the phases of the \ac{RIS}
elements. In \cite{perovic2021achievable}, the \ac{PGM} was used
to maximize the achievable rate in an \ac{RIS}-based \ac{MIMO} communication
system.

On the other hand, \ac{SM}, or more generally \ac{IM}, has been
widely under investigation in the last decade due to its inherent
energy efficiency. Massive connectivity results in enormously increasing
energy consumption, while in \ac{IM} part of the information is conveyed
by the indices of the available resources, e.g. transmit or receive
antennas, frequency-domain subcarriers, etc., such that only a subset
of the energy-consuming resources are activated at any time; this
characterizes \ac{IM} as an energy-efficient solution. Therefore,
\ac{IM} has been recognized as another promising technology in \ac{6G}
systems, thus motivating researchers to develop \ac{RIS}-based \ac{IM}
systems. In particular, \acf{RIS-SSK} and \ac{RIS-SM} systems were
introduced in \cite{basar2020reconfigurable}, where indices of \emph{receive}
antennas can be considered to realize \ac{IM}, while an \ac{RIS}
was deployed at the transmitter as an access point. The \ac{BER}
performance of these systems was investigated and compared to that
of the \ac{RIS-AP} scheme in \cite{basar2019transmission}. Inspired
by \cite{basar2020reconfigurable}, the authors in \cite{canbilen2020reconfigurable}
analyzed the \ac{BER} performance of a \ac{MISO} \ac{RIS}-assisted
\ac{SSK} system where the index of the \emph{transmit} antenna conveyed
the information. In \cite{ma2020large}, a new \ac{RIS}-based \ac{SM}
paradigm was proposed in which the indices of both the transmit and
receive antennas are selected in order to convey information; hence,
the spectral efficiency is increased at the expense of a higher receiver
complexity. Inspired by \acfi{QSM} proposed in \cite{mesleh2014quadrature},
which implements \ac{SM} independently on the real (in-phase) and
imaginary (quadrature) dimensions, \ac{RIS-RQRM} was proposed in
\cite{yuan2021receive}. In this approach, the \ac{RIS} elements
are divided into two equal groups each of which independently performs
\ac{SM}. Although this technique doubles the spectral efficiency,
it suffers from a degraded \ac{BER} performance compared to \ac{RIS-SM}.
The concept of \ac{SM} has been extended to \ac{GSM} in \cite{albinsaid2021multiple}
and \cite{zhang2021large} to increase the spectral efficiency of
the \ac{RIS}-based wireless system. However, in these schemes, the
\ac{RIS} elements are divided into a number of groups in order to
maintain beamforming toward multiple selected receive antennas, which
yields a reduction in the received signal power. Another \ac{RIS}-based
\ac{SSK} scheme (\ac{SSK} at the transmitter side, similar to \cite{canbilen2020reconfigurable})
was proposed in \cite{li2021space}. In this scheme, however, the
\ac{RIS} was assumed to have no knowledge regarding the transmit
data, and therefore an optimization problem was defined to maximize
the minimum squared channel-imprinted Euclidean distance at the receiver.
The authors also proposed an extension to the scheme such that the
\ac{RIS}, in addition to reflecting the incident \ac{SSK} signal,
also transmits its own information via an Alamouti \ac{STBC}. In
\cite{lin2021reconfigurable} and \cite{lin2021reconfigurable_journal},
the concept of \ac{IM} has been applied within the \ac{RIS} entity,
in which the \ac{RIS} elements are divided into two groups so that
\ac{IM} can be implemented on these groups in order to transmit environmental
data to the receiver, similar to the \acf{RPM} scheme of \cite{lin2020reconfigurable}.

However, in all of the aforementioned studies, each group of \ac{RIS}
elements \emph{separately} targets one receive or transmit antenna
for the purpose of implementing \ac{SM}. Despite the advantages of
\ac{SM}, the spectral efficiency of \ac{SM} needs to be improved,
and this can be performed by introducing other variants of \ac{SM}
such as \ac{QSM} or \ac{GSM}, which require two or more antennas
to be activated. The proposed solutions so far, however, cut down
the effective number of \ac{RIS} elements. Hence, the spectral efficiency
increases only at the cost of a reduction in the received signal power.

Against this background, in this paper we introduce a new paradigm
for \ac{RIS}-based \ac{IM} in which the information is conveyed
through the indices of two selected receive antennas. The resulting
approach has the property that \emph{all} \ac{RIS} elements can
\emph{independently} perform beamforming onto the two selected receive
antennas. The contributions of this paper are as follows: 
\begin{itemize}
\item Inspired by \ac{QSM}, we propose a novel \ac{RIS}-assisted \ac{IM}
scheme, namely \ac{RIS-RQSSK}, in which all \ac{RIS} elements simultaneously
maximize the \ac{SNR} of the in-phase and quadrature components of
the received signal at the selected antennas. That is, the \ac{SNR}
associated to the real part of the signal at one antenna and the \ac{SNR}
associated to the imaginary part of the signal at the second antenna
are maximized in order to be detectable by a simple \ac{GD}. Therefore,
the spectral efficiency is increased compared to conventional \ac{SSK},
without significant additional complexity or cost.
\item We propose a max-min optimization problem in order to maximize the
two relevant \ac{SNR} components. Since this problem is non-convex,
we determine its dual problem, which is convex and admits an analytical
solution. Specifically, the joint optimization of the \ac{RIS} phase
shifts reduces to solving a single-variable equation in order to determine
each optimal \ac{RIS} phase shift. We also show that with a large
number of \ac{RIS} elements the solution of this equation tends to
a constant value, thus providing a very simple design procedure to
control the phase of the \ac{RIS} elements.
\item We analyze the \ac{ABEP} of the \ac{RIS-RQSSK} system. We also use
approximations in order to derive a closed-form approximate \ac{ABEP}
which is tight in the \ac{SNR} range of interest for a large number
of \ac{RIS} elements.
\item Finally, we investigate the \ac{BER} performance of the \ac{RIS-RQSSK}
system through numerical simulations and compare the results with
those of the most prominent recently proposed schemes. The results
show that the proposed \ac{RIS-RQSSK} system significantly outperforms
these benchmark schemes, and that the performance enhancement improves
with an increasing number of receive antennas. 
\end{itemize}
The rest of this paper is organized as follows. We describe the \ac{RIS-RQSSK}
system model in Section~\ref{sec:System-Model}. In Section~\ref{sec:Problem-Formulation},
we formulate the optimization problem and investigate its analytical
solution. The \ac{ABEP} performance of the proposed \ac{RIS-RQSSK}
system with and without polarity bits is analyzed in Section~\ref{sec:Performance-Analysis}.
Numerical simulations and comparisons with the benchmark schemes are
provided in Section~\ref{sec:Numerical-Results}. Finally, Section~\ref{sec:Conclusion}
concludes this paper.

\emph{Notation:} Boldface lower-case letters denote column vectors,
and boldface upper-case letters denote matrices. $\mathbf{v}>0$ (resp.,
$\mathbf{v}\geq0$) indicates that all elements in vector $\mathbf{v}$
are positive (resp., non-negative). $\mathbf{u}\odot\mathbf{v}$ represents
the element-wise product of two equal-sized vectors $\mathbf{u}$
and $\mathbf{v}$. The superscripts $\left(\cdot\right)^{T}$ and
$\left(\cdot\right)^{H}$ denote transpose and Hermitian transpose,
respectively. $\left(\cdot\right)^{\mathcal{R}}$ and $\left(\cdot\right)^{\mathcal{I}}$
denote the real and imaginary components of a scalar/vector, respectively.
$\mathsf{\mathbb{E}}\left\{ \cdot\right\} $ and $\mathbb{V}\left\{ \cdot\right\} $,
respectively, denote the expectation and variance operator. $\mathcal{N}\left(\mu,\sigma^{2}\right)$
(resp., $\mathcal{CN}\left(\mu,\sigma^{2}\right)$) represents the
normal (resp., complex normal) distribution with mean $\mu$ and variance
$\sigma^{2}$. $\chi_{d}^{2}$ and $\chi_{d}^{2}\left(\lambda\right)$
denote central and non-central chi-square distributions, respectively,
with $d$ degrees of freedom and non-centrality parameter $\lambda$.
Finally, the set of complex matrices of size $m\times n$ is denoted
by $\mathbb{C}^{m\times n}$.

\section{System Model\label{sec:System-Model}}

\begin{figure}[t]
\begin{centering}
\includegraphics[scale=0.23]{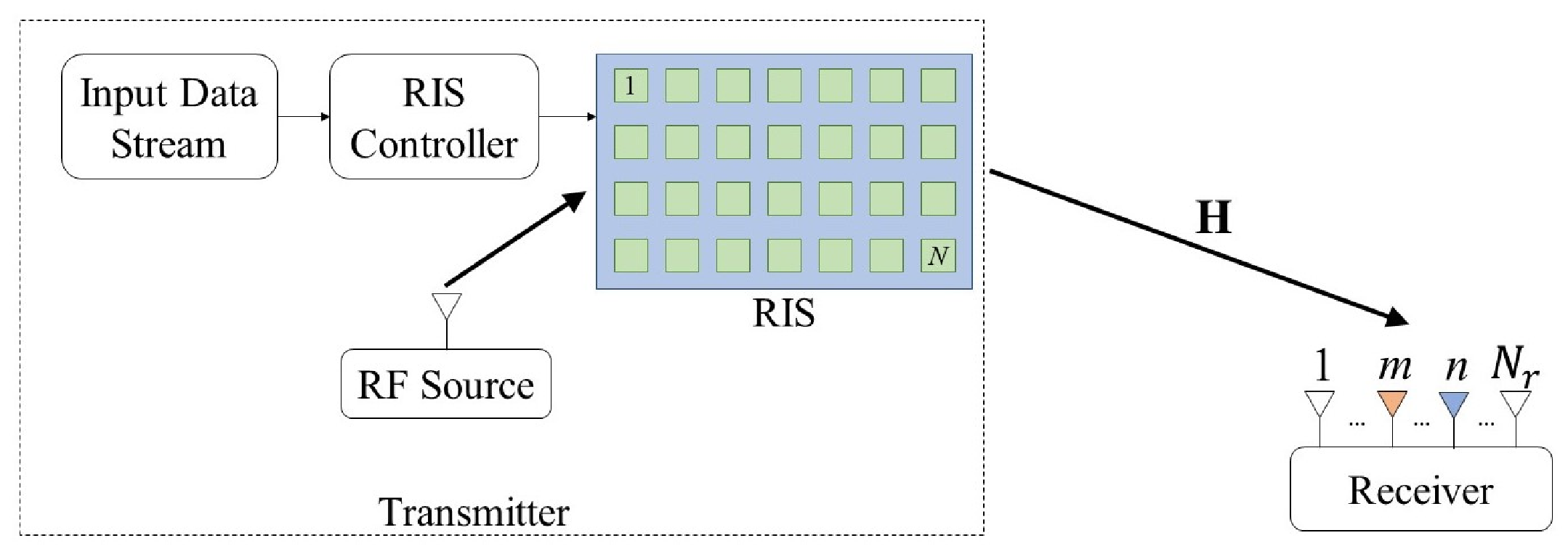}
\par\end{centering}
\caption{A schematic representation of the proposed \ac{RIS-RQSSK} system.\label{fig:schematic}}
\end{figure}

The proposed \acf{RIS-RQSSK} system is illustrated in Fig.~\ref{fig:schematic}.
In this scenario, we consider an \ac{RIS}-AP\footnote{It is also possible to use a phased array antenna to implement the
proposed RQSSK system; however, RIS is a very promising new technology
which has many advantages over existing technologies (nearly-passive
operation, full-duplex capability without significant self-interference,
etc.), and is being considered as a potential candidate for future
smart radio networks. For this reason, we investigate the use of the
RIS-AP in this research work.} scheme in which the \ac{RIS} forms part of the transmitter and it
reflects the incident wave emitted from a single RF source which is
located in the vicinity of the \ac{RIS} such that the path loss of
the link between the \ac{RIS} and the RF source is negligible. The
\ac{RIS} is equipped with $N$ reflecting elements whose phases are
controlled by the transmitter through the \ac{RIS} controller. We
assume that the receiver, which is equipped with $N_{r}$ antennas,
can only receive the signal reflected from the \ac{RIS} elements\footnote{Note that this is not a naive assumption; if a direct path from the
RF source to destination exists, this is mathematically equivalent
to the addition of another RIS element, in the sense that the channel
model is still given by an expression of the form of (\ref{eq:system model y}).}. The concept of RIS-AP was first introduced in \cite{basar2019wireless}
and \cite{basar2019transmission}; later, in \cite{basar2020reconfigurable},
this model was extended to cover \ac{RIS}-aided wireless communication
systems using \ac{IM} techniques.

The baseband receive signal at receive antenna $l$ is given by
\begin{equation}
y_{l}=\sqrt{E_{s}}\mathbf{h}_{l}\boldsymbol{\theta}+n_{l},\label{eq:system model y}
\end{equation}
where $\mathbf{h}_{l}=\left[h_{l,1},h_{l,2},\dots,h_{l,N}\right]\in\mathbb{C}^{1\times N}$
is the $l$-th row of $\mathbf{H}\in\mathbb{C}^{N_{r}\times N}$,
which is the channel matrix of the link between the \ac{RIS} and
the receiver whose elements are i.i.d and distributed according to
$\mathcal{CN}\left(0,1\right)$. $\boldsymbol{\theta}\in\mathbb{C}^{N\times1}$
is the vector that consists of the reflection coefficients of the
\ac{RIS} elements, such that $\left|\theta_{i}\right|=1$ for $i=1,2,\dots,N$
(here we assume lossless reflection from the \ac{RIS}). $E_{s}$
is the transmitted energy from the RF source per \ac{IM} symbol,
and $n_{l}\in\mathbb{C}$ is the additive white Gaussian noise at
the $l$-th receive antenna that is distributed according to $\mathcal{CN}\left(0,N_{0}\right)$.
Hence, the \ac{SNR} is equal to $E_{s}/N_{0}$.

In the proposed RIS-RQSSK system, the input data bit stream is split
into blocks of $R=2\left(\log_{2}N_{r}+1\right)$ bits; one packet
of $\log_{2}N_{r}+1$ bits is mapped to an in-phase (real) signal,
while the other packet of $\log_{2}N_{r}+1$ bits is mapped to a quadrature
(imaginary) signal. In each of the two constituent packets of $(\log_{2}N_{r}+1)$
bits, the first $\log_{2}N_{r}$ bits are mapped to one receive antenna
index and the final bit determines the corresponding polarity. In
fact, the transmitted data bits determine the indices of \emph{two}
selected receive antennas. Unlike \ac{RIS-SSK} \cite{basar2020reconfigurable},
where the transmitter aims to maximize the \ac{SNR} at one specific
receive antenna, in the proposed scheme the transmitter independently
selects two receive antennas. That is, the transmitter aims to simultaneously
maximize the \ac{SNR} of the real part of the signal at the first
selected receive antenna, while also maximizing the \ac{SNR} of the
imaginary part of the signal at the second selected receive antenna.
Recalling RSM techniques within MIMO systems, receive antenna selection
can be performed via implementing a precoding matrix along with a
transmit vector at the transmitter; however, in the RIS-AP scheme,
where the transmitter (RF source) is equipped with only one antenna,
passive beamforming via adjusting the phase of the RIS elements conducts
the antenna selection task. In the next section, we will show in detail
how the RIS reflection coefficient vector $\boldsymbol{\theta}$ is
optimized to perform the so-called passive beamforming. Note that
in this work, we assume that the transmitter has perfect knowledge
of the \ac{CSI} that is needed in order to calculate the optimal
RIS phase shifts.

Suppose that $m$ and $n$ are the selected antenna indices for the
real and imaginary parts, respectively. After expanding (\ref{eq:system model y})
for selected antennas $m$ and $n$, and separating the real and imaginary
parts, we have 
\begin{align}
y_{m}^{\mathcal{R}} & =\sqrt{E_{s}}\left[\mathbf{h}_{m}^{\mathcal{R}}\boldsymbol{\theta}^{\mathcal{R}}-\mathbf{h}_{m}^{\mathcal{I}}\boldsymbol{\theta}^{\mathcal{I}}\right]+n_{m}^{\mathcal{R}},\label{eq:real y}
\end{align}
\begin{align}
y_{n}^{\mathcal{I}} & =\sqrt{E_{s}}\left[\mathbf{h}_{n}^{\mathcal{R}}\boldsymbol{\theta}^{\mathcal{I}}+\mathbf{h}_{n}^{\mathcal{I}}\boldsymbol{\theta}^{\mathcal{R}}\right]+n_{n}^{\mathcal{I}}.\label{eq:imag y}
\end{align}
The receive signal components (\ref{eq:real y}) and (\ref{eq:imag y})
suggest that $\boldsymbol{\theta}$ needs to be optimized in order
to maximize the relevant \acp{SNR}. After this \ac{SNR} maximization
has been performed, a simple but effective \acfi{GD} can be employed
to detect the selected receive antennas without the need for any knowledge
of the \ac{CSI} at the receiver. Then, the \ac{GD} operates via
\begin{align}
\hat{m} & =\arg\max_{m\in\{1,2,\dots,N_{r}\}}\left\{ \left(y_{m}^{\mathcal{R}}\right)^{2}\right\} ,\label{eq:GD m_hat}\\
\hat{n} & =\arg\max_{n\in\{1,2,\dots,N_{r}\}}\left\{ \left(y_{n}^{\mathcal{I}}\right)^{2}\right\} .\label{eq:GD n_hat}
\end{align}
That is, the \ac{GD} estimates the antenna indices by independently
searching over the instantaneous energy of the real and imaginary
parts of the signal at the receive antennas and choosing the one with
highest energy in each case (note that we may have $\hat{m}=\hat{n}$).
After this, the polarity bits can be detected simply by testing the
sign of each of the values $y_{\hat{m}}^{\mathcal{R}}$ and $y_{\hat{n}}^{\mathcal{I}}$.

While the GD, being a low-complexity and energy-efficient detector,
is considered as the superior approach for symbol detection at the
user side, optimum maximum likelihood (ML) detection can also be implemented
at the receiver. The ML detector operates via
\begin{align}
 & \left(\hat{m},\hat{n},\hat{d}_{m},\hat{d}_{n}\right)=\nonumber \\
 & \arg\min_{m,n,d_{m},d_{n}}\sum_{l=1}^{N_{r}}\left(y_{l}-\sqrt{E_{s}}\mathbf{h}_{l}\boldsymbol{\theta}^{\star}\left(m,n,d_{m},d_{n}\right)\right)^{2},\label{eq:ML}
\end{align}
where $\boldsymbol{\theta}^{\star}\left(m,n,d_{m},d_{n}\right)$ is
the vector of optimum phase shifts of the RIS elements corresponding
to both the selected pair of antennas $\left(m,n\right)$ and the
pair of polarity bits $\left(d_{m},d_{n}\right)$. In the next section
we will show how $\boldsymbol{\theta}^{\star}$ is related to $\left(m,n,d_{m},d_{n}\right)$.
We will see later in Section \ref{sec:Numerical-Results} that the
performance of the ML detector is considerably close to that of the
GD, especially for large values of $N$, so that the additional complexity
of ML is not worthy of being implemented.

\section{Problem Formulation\label{sec:Problem-Formulation}}

In this section, we define the optimization problem for the proposed
\ac{RIS-RQSSK} system to find the optimum phase shifts of the \ac{RIS}
elements. Here $m$ and $n$ denote the selected antenna indices based
on the input data bits for the real and imaginary parts, respectively
(note that we may have $m=n$). Therefore, the transmitter aims to
maximize the \ac{SNR} of the real part of the signal at antenna $m$,
denoted by $\mathrm{SNR}_{m}^{(R)}$, and the \ac{SNR} of the imaginary
part of the signal at antenna $n$, denoted by $\mathrm{SNR}_{n}^{(I)}$,
at the same time. Recalling (\ref{eq:real y}) and (\ref{eq:imag y}),
$\mathrm{SNR}_{m}^{(R)}$ and $\mathrm{SNR}_{n}^{(I)}$ may be expressed
as 
\begin{equation}
\mathrm{SNR}_{m}^{(R)}=\frac{2E_{s}\left(\mathbf{h}_{m}^{\mathcal{R}}\boldsymbol{\theta}^{\mathcal{R}}-\mathbf{h}_{m}^{\mathcal{I}}\boldsymbol{\theta}^{\mathcal{I}}\right)^{2}}{N_{0}},\label{eq:SNR m}
\end{equation}
\begin{equation}
\mathrm{SNR}_{n}^{(I)}=\frac{2E_{s}\left(\mathbf{h}_{n}^{\mathcal{R}}\boldsymbol{\theta}^{\mathcal{I}}+\mathbf{h}_{n}^{\mathcal{I}}\boldsymbol{\theta}^{\mathcal{R}}\right)^{2}}{N_{0}}.\label{eq:SNR n}
\end{equation}
Note that the same variables exist in (\ref{eq:SNR m}) and (\ref{eq:SNR n}),
therefore, it is not feasible to separately maximize these \ac{SNR}
values. Hence, we define a max-min optimization problem to maximize
the minimum of these two \acp{SNR}. This optimization problem can
be defined as\footnote{It is worth mentioning that \emph{minimizing} the relevant in-phase/quadrature
received energy at the \emph{non-selected} antennas is also desired;
however, this is not straightforward to achieve via simply adjusting
the phase shifts of the RIS elements, i.e., passive beamforming. Therefore,
here we only target the maximization of the received energy at the
selected antennas. In addition, an insight into the performance of
the system can be obtained when our proposed approach is used, namely,
we will see later in Theorems 1-3 that our proposed solution provides
the maximum average signal amplitudes (positive or negative, depending
on the polarity bits) at the selected antennas while maintaining an
average signal amplitude of zero at the non-selected antennas.} 
\begin{align}
\underset{\boldsymbol{\theta}^{\mathcal{R}},\boldsymbol{\theta}^{\mathcal{I}}}{\max} & \ \min\left(\left|\mathbf{h}_{m}^{\mathcal{R}}\boldsymbol{\theta}^{\mathcal{R}}-\mathbf{h}_{m}^{\mathcal{I}}\boldsymbol{\theta}^{\mathcal{I}}\right|,\left|\mathbf{h}_{n}^{\mathcal{R}}\boldsymbol{\theta}^{\mathcal{I}}+\mathbf{h}_{n}^{\mathcal{I}}\boldsymbol{\theta}^{\mathcal{R}}\right|\right)\label{eq: MAX op problem}\\
\mbox{s.t.}\;\; & \ \left(\theta_{i}^{\mathcal{R}}\right)^{2}+\left(\theta_{i}^{\mathcal{I}}\right)^{2}=1,\ \mbox{for all}\;i=1,2,\dots,N.\nonumber 
\end{align}
Then, re-expressing in the standard form by defining an auxiliary
parameter $t$, the optimization problem can be defined as

\begin{align}
\underset{\boldsymbol{\theta}^{\mathcal{R}},\boldsymbol{\theta}^{\mathcal{I}},t}{\mathrm{\min}} & \;f_{0}\left(\boldsymbol{\theta}^{\mathcal{R}},\boldsymbol{\theta}^{\mathcal{I}},t\right)\triangleq t\label{eq:STANDARD op problem}\\
\mbox{s.t.}\quad & \;f_{1}\left(\boldsymbol{\theta}^{\mathcal{R}},\boldsymbol{\theta}^{\mathcal{I}},t\right)\triangleq-\left|\mathbf{h}_{m}^{\mathcal{R}}\boldsymbol{\theta}^{\mathcal{R}}-\mathbf{h}_{m}^{\mathcal{I}}\boldsymbol{\theta}^{\mathcal{I}}\right|-t\leq0,\nonumber \\
 & \;f_{2}\left(\boldsymbol{\theta}^{\mathcal{R}},\boldsymbol{\theta}^{\mathcal{I}},t\right)\triangleq-\left|\mathbf{h}_{n}^{\mathcal{R}}\boldsymbol{\theta}^{\mathcal{I}}+\mathbf{h}_{n}^{\mathcal{I}}\boldsymbol{\theta}^{\mathcal{R}}\right|-t\leq0,\nonumber \\
 & \;h_{i}\left(\boldsymbol{\theta}^{\mathcal{R}},\boldsymbol{\theta}^{\mathcal{I}},t\right)\triangleq\left(\theta_{i}^{\mathcal{R}}\right)^{2}+\left(\theta_{i}^{\mathcal{I}}\right)^{2}-1=0,\nonumber \\
 & \;\forall i=1,2,\dots,N.\nonumber 
\end{align}
This problem is a non-convex optimization problem, as there exist
non-linear equality constraints. Hence, to reformulate this problem
in the form of a convex optimization problem, we consider the Lagrange
dual of this problem. Then, we investigate the analytical solution
of the resulting convex problem. Moreover, it is worth noting that
due to the appearance of the absolute value operation in the definitions
of $f_{1}$ and $f_{2}$, the Lagrange dual function needs to be calculated
in four cases:
\begin{casenv}
\item 
\[
\mathbf{h}_{m}^{\mathcal{R}}\boldsymbol{\theta}^{\mathcal{R}}-\mathbf{h}_{m}^{\mathcal{I}}\boldsymbol{\theta}^{\mathcal{I}}\geq0\;\mbox{and}\;\mathbf{h}_{n}^{\mathcal{R}}\boldsymbol{\theta}^{\mathcal{I}}+\mathbf{h}_{n}^{\mathcal{I}}\boldsymbol{\theta}^{\mathcal{R}}\geq0,
\]
\item 
\[
\mathbf{h}_{m}^{\mathcal{R}}\boldsymbol{\theta}^{\mathcal{R}}-\mathbf{h}_{m}^{\mathcal{I}}\boldsymbol{\theta}^{\mathcal{I}}\geq0\;\mbox{and}\;\mathbf{h}_{n}^{\mathcal{R}}\boldsymbol{\theta}^{\mathcal{I}}+\mathbf{h}_{n}^{\mathcal{I}}\boldsymbol{\theta}^{\mathcal{R}}<0,
\]
\item 
\[
\mathbf{h}_{m}^{\mathcal{R}}\boldsymbol{\theta}^{\mathcal{R}}-\mathbf{h}_{m}^{\mathcal{I}}\boldsymbol{\theta}^{\mathcal{I}}<0\;\mbox{and}\;\mathbf{h}_{n}^{\mathcal{R}}\boldsymbol{\theta}^{\mathcal{I}}+\mathbf{h}_{n}^{\mathcal{I}}\boldsymbol{\theta}^{\mathcal{R}}\geq0,
\]
\item 
\[
\mathbf{h}_{m}^{\mathcal{R}}\boldsymbol{\theta}^{\mathcal{R}}-\mathbf{h}_{m}^{\mathcal{I}}\boldsymbol{\theta}^{\mathcal{I}}<0\;\mbox{and}\;\mathbf{h}_{n}^{\mathcal{R}}\boldsymbol{\theta}^{\mathcal{I}}+\mathbf{h}_{n}^{\mathcal{I}}\boldsymbol{\theta}^{\mathcal{R}}<0,
\]
\end{casenv}
where each case corresponds to a particular combination of the polarity
bits, and it is then required to solve for the optimum phase angles.
In the following, we show in detail how this problem can be solved
for Case 1; the solution for the other three cases proceeds similarly.
In Case 1, the Lagrange function associated with the problem in (\ref{eq:STANDARD op problem})
is defined as \cite{boyd2004convex}{\small{}
\begin{align}
 & L\left(\boldsymbol{\theta}^{\mathcal{R}},\boldsymbol{\theta}^{\mathcal{I}},t,\boldsymbol{\lambda},\boldsymbol{\nu}\right)=\nonumber \\
 & f_{0}\left(\boldsymbol{\theta}^{\mathcal{R}},\boldsymbol{\theta}^{\mathcal{I}},t\right)+\sum_{j=1}^{2}\lambda_{j}f_{j}\left(\boldsymbol{\theta}^{\mathcal{R}},\boldsymbol{\theta}^{\mathcal{I}},t\right)+\sum_{i=1}^{N}\nu_{i}h_{i}\left(\boldsymbol{\theta}^{\mathcal{R}},\boldsymbol{\theta}^{\mathcal{I}},t\right)=\nonumber \\
 & \left(1-\lambda_{1}-\lambda_{2}\right)t-\lambda_{1}\left(\mathbf{h}_{m}^{\mathcal{R}}\boldsymbol{\theta}^{\mathcal{R}}-\mathbf{h}_{m}^{\mathcal{I}}\boldsymbol{\theta}^{\mathcal{I}}\right)-\lambda_{2}\left(\mathbf{h}_{n}^{\mathcal{R}}\boldsymbol{\theta}^{\mathcal{I}}+\mathbf{h}_{n}^{\mathcal{I}}\boldsymbol{\theta}^{\mathcal{R}}\right)\nonumber \\
 & +\sum_{i=1}^{N}\nu_{i}\left(\left(\theta_{i}^{\mathcal{R}}\right)^{2}+\left(\theta_{i}^{\mathcal{I}}\right)^{2}-1\right),\label{eq:Lagrange function}
\end{align}
}where $\boldsymbol{\lambda}=\left[\lambda_{1},\lambda_{2}\right]^{T}\geq0$
and $\boldsymbol{\nu}=\left[\nu_{1},\nu_{2},\dots,\nu_{N}\right]^{T}$
are vectors of Lagrange multipliers. Considering the Lagrange function,
the objective function of the Lagrange dual problem is computed as 

\begin{equation}
g\left(\boldsymbol{\lambda},\boldsymbol{\nu}\right)=\inf_{\boldsymbol{\theta}^{\mathcal{R}},\boldsymbol{\theta}^{\mathcal{I}},t}L\left(\boldsymbol{\theta}^{\mathcal{R}},\boldsymbol{\theta}^{\mathcal{I}},t,\boldsymbol{\lambda},\boldsymbol{\nu}\right).\label{eq:Dual definition}
\end{equation}
The Lagrange function in (\ref{eq:Lagrange function}) is a quadratic
function of $\bigl(\boldsymbol{\theta}^{\mathcal{R}},\boldsymbol{\theta}^{\mathcal{I}}\bigr)$,
therefore, it is lower bounded if $\boldsymbol{\nu}>0$, i.e., if
the function is convex quadratic in $\bigl(\boldsymbol{\theta}^{\mathcal{R}},\boldsymbol{\theta}^{\mathcal{I}}\bigr)$.
Therefore, we can find the minimizing $\bigl(\boldsymbol{\theta}^{\mathcal{R}},\boldsymbol{\theta}^{\mathcal{I}}\bigr)$
from the optimality conditions

\begin{align}
 & \nabla_{\boldsymbol{\theta}^{\mathcal{R}}}L\left(\boldsymbol{\theta}^{\mathcal{R}},\boldsymbol{\theta}^{\mathcal{I}},t,\boldsymbol{\lambda},\boldsymbol{\nu}\right)=\nonumber \\
 & -\lambda_{1}\left(\mathbf{h}_{m}^{\mathcal{R}}\right)^{T}-\lambda_{2}\left(\mathbf{h}_{n}^{\mathcal{I}}\right)^{T}+2\boldsymbol{\nu}\odot\boldsymbol{\theta}^{\mathcal{R}}=0,\label{eq:Gradient of L}
\end{align}
and
\begin{align}
 & \nabla_{\boldsymbol{\theta}^{\mathcal{I}}}L\left(\boldsymbol{\theta}^{\mathcal{R}},\boldsymbol{\theta}^{\mathcal{I}},t,\boldsymbol{\lambda},\boldsymbol{\nu}\right)=\nonumber \\
 & \lambda_{1}\left(\mathbf{h}_{m}^{\mathcal{I}}\right)^{T}-\lambda_{2}\left(\mathbf{h}_{n}^{\mathcal{R}}\right)^{T}+2\boldsymbol{\nu}\odot\boldsymbol{\theta}^{\mathcal{I}}=0.\label{eq:Gradient of L-2}
\end{align}
Thus, we obtain 
\begin{align}
\theta_{i}^{\mathcal{R}^{\star}} & =\frac{\lambda_{1}h_{m,i}^{\mathcal{R}}+\lambda_{2}h_{n,i}^{\mathcal{I}}}{2\nu_{i}},\ i=1,2,\dots,N,\label{eq:optimum theta_R}\\
\theta_{i}^{\mathcal{I}^{\star}} & =\frac{-\lambda_{1}h_{m,i}^{\mathcal{I}}+\lambda_{2}h_{n,i}^{\mathcal{R}}}{2\nu_{i}},\ i=1,2,\dots,N.\label{eq:optimum theta_I}
\end{align}
In addition, $L\left(\boldsymbol{\theta}^{\mathcal{R}},\boldsymbol{\theta}^{\mathcal{I}},t,\boldsymbol{\lambda},\boldsymbol{\nu}\right)$
is a linear function of $t$; therefore, it is bounded below only
when the coefficient of $t$ is equal to zero, i.e.,
\begin{equation}
1-\lambda_{1}-\lambda_{2}=0.\label{eq:=00005Clambdacondition}
\end{equation}
Then, by substituting (\ref{eq:optimum theta_R}), (\ref{eq:optimum theta_I})
and (\ref{eq:=00005Clambdacondition}) into (\ref{eq:Lagrange function}),
the Lagrange dual can be written as (\ref{eq:DUAL lagrange function})
shown on the next page.
\begin{figure*}[tbh]
\begin{align}
 & g\left(\boldsymbol{\lambda},\boldsymbol{\nu}\right)=\begin{cases}
-\frac{1}{4}\sum_{i=1}^{N}\frac{1}{\nu_{i}}\left[\left(\lambda_{1}h_{m,i}^{\mathcal{R}}+\lambda_{2}h_{n,i}^{\mathcal{I}}\right)^{2}+\left(-\lambda_{1}h_{m,i}^{\mathcal{I}}+\lambda_{2}h_{n,i}^{\mathcal{R}}\right)^{2}\right]-\sum_{i=1}^{N}\nu_{i}, & \begin{array}{l}
\lambda_{1}+\lambda_{2}=1,\boldsymbol{\lambda}\geq0,\boldsymbol{\nu}>0,\end{array}\\
-\infty, & \mbox{otherwise.}
\end{cases}\label{eq:DUAL lagrange function}\\
\hline \nonumber 
\end{align}
\end{figure*}
 As a result, the Lagrange dual problem is defined as
\begin{eqnarray}
\underset{\boldsymbol{\lambda},\boldsymbol{\nu}}{\mathrm{\min}} &  & -g\left(\boldsymbol{\lambda},\boldsymbol{\nu}\right)\label{eq:Dual problem}\\
\mbox{s.t.} &  & \lambda_{1}+\lambda_{2}-1=0\nonumber \\
 &  & \lambda_{1},\lambda_{2}\geq0\nonumber \\
 &  & \nu_{i}>0,\ i=1,2,\dots,N.\nonumber 
\end{eqnarray}
Note that $g\left(\boldsymbol{\lambda},\boldsymbol{\nu}\right)$ is
a concave function of $\boldsymbol{\nu}$ with $\boldsymbol{\nu}>0$,
thus, the optimal point over $\boldsymbol{\nu}$ can be found from
the optimality condition $\nabla_{\boldsymbol{\nu}}g\left(\boldsymbol{\lambda},\boldsymbol{\nu}\right)=0$,
which yields 
\begin{equation}
\nu_{i}^{\star}=\frac{1}{2}\sqrt{\left(\lambda_{1}h_{m,i}^{\mathcal{R}}+\lambda_{2}h_{n,i}^{\mathcal{I}}\right)^{2}+\left(-\lambda_{1}h_{m,i}^{\mathcal{I}}+\lambda_{2}h_{n,i}^{\mathcal{R}}\right)^{2}},\label{eq:optimal nu}
\end{equation}
for all $i=1,2,\dots,N$. Substituting (\ref{eq:optimal nu}) into
(\ref{eq:DUAL lagrange function}), the problem (\ref{eq:Dual problem})
is updated as
\begin{align}
\underset{\boldsymbol{\lambda}}{\mathrm{\min}} & \ \sum_{i=1}^{N}\sqrt{\left(\lambda_{1}A_{i}+\lambda_{2}B_{i}\right)^{2}+\left(\lambda_{1}C_{i}+\lambda_{2}D_{i}\right)^{2}}\label{eq:Dual problem-lambda}\\
\mbox{s.t.} & \ \lambda_{1},\lambda_{2}\geq0,\nonumber \\
 & \ \lambda_{1}+\lambda_{2}-1=0,\nonumber 
\end{align}
where we define $A_{i}=h_{m,i}^{\mathcal{R}}$, $B_{i}=h_{n,i}^{\mathcal{I}}$,
$C_{i}=-h_{m,i}^{\mathcal{I}}$ and $D_{i}=h_{n,i}^{\mathcal{R}}$,
to simplify the notation. $\lambda_{2}$ also can be eliminated, hence
the problem can be expressed as
\begin{align}
\underset{\lambda_{1}}{\min} & \ \sum_{i=1}^{N}\sqrt{\left(\lambda_{1}A_{i}+\left(1-\lambda_{1}\right)B_{i}\right)^{2}+\left(\lambda_{1}C_{i}+\left(1-\lambda_{1}\right)D_{i}\right)^{2}}\label{eq:Dual problem-lambda-1}\\
\mbox{s.t.} & \ 0\leq\lambda_{1}\leq1.\nonumber 
\end{align}
The Lagrange function associated with this problem is given by 
\begin{align}
 & L^{(1)}\left(\lambda_{1},\alpha,\beta\right)=\nonumber \\
 & \sum_{i=1}^{N}\sqrt{\left(\lambda_{1}A_{i}+\left(1-\lambda_{1}\right)B_{i}\right)^{2}+\left(\lambda_{1}C_{i}+\left(1-\lambda_{1}\right)D_{i}\right)^{2}}\nonumber \\
 & -\alpha\lambda_{1}+\beta\left(\lambda_{1}-1\right),\label{eq:Lagrange(1) function}
\end{align}
where $\alpha\geq0$ and $\beta\geq0$ are the Lagrange multipliers
for this problem. We know that for a convex problem, any points that
satisfy the \acfi{KKT} conditions \cite{boyd2004convex} are both
primal and dual optimal. Hence, we write the \ac{KKT} conditions
for the problem (\ref{eq:Dual problem-lambda-1}) with respect to
the primal and dual optimal points $\left(\lambda_{1}^{\star},\alpha^{\star},\beta^{\star}\right)$
as in (\ref{eq:KKT conditions}) shown on the next page,
\begin{figure*}[tbh]
\begin{align}
1.\;\lambda_{1}^{\star}\geq0\ ;\ 2.\;\lambda_{1}^{\star}-1\leq0\ ;\ 3.\;\alpha^{\star}\geq0\ ;\ 4.\;\beta^{\star}\geq0\ ;\ 5.\;\alpha^{\star}\lambda_{1}^{\star}=0\ ;\ 6.\;\beta^{\star}\left(\lambda_{1}^{\star}-1\right)=0\ ;\nonumber \\
7.\;\sum_{i=1}^{N}\frac{\left(A_{i}-B_{i}\right)\left(\lambda_{1}^{\star}A_{i}+\left(1-\lambda_{1}^{\star}\right)B_{i}\right)+\left(C_{i}-D_{i}\right)\left(\lambda_{1}^{\star}C_{i}+\left(1-\lambda_{1}^{\star}\right)D_{i}\right)}{\sqrt{\left(\lambda_{1}^{\star}A_{i}+\left(1-\lambda_{1}^{\star}\right)B_{i}\right)^{2}+\left(\lambda_{1}^{\star}C_{i}+\left(1-\lambda_{1}^{\star}\right)D_{i}\right)^{2}}}-\alpha^{\star}+\beta^{\star}=0;\label{eq:KKT conditions}\\
\hline \nonumber 
\end{align}
\end{figure*}
 where the \ac{KKT} condition 7 is equivalent to $\nabla_{\lambda_{1}}L^{(1)}\left(\lambda_{1},\alpha,\beta\right)=0$;
therefore, the resulting $\lambda_{1}^{\star}$ minimizes $L^{(1)}\left(\lambda_{1},\alpha^{\star},\beta^{\star}\right)$
over $\lambda_{1}$. We can now directly solve these equations to
find $\left(\lambda_{1}^{\star},\alpha^{\star},\beta^{\star}\right)$.

To analyze \ac{KKT} conditions 5 and 6 in (\ref{eq:KKT conditions}),
we consider four possible cases: 1) $\alpha^{\star}=0$, $\beta^{\star}\neq0$,
2) $\alpha^{\star}\neq0$, $\beta^{\star}=0$, 3) $\alpha^{\star}=0$,
$\beta^{\star}=0$ and 4) $\alpha^{\star}\neq0$, $\beta^{\star}\neq0$.
It is easy to see that the final case is not feasible; therefore,
we investigate the other three cases, as follows:

1) $\alpha^{\star}=0$, $\beta^{\star}\neq0$ ; In this case, from
\ac{KKT} condition 6 in (\ref{eq:KKT conditions}) we find $\lambda_{1}^{\star}=1$.
Since $\beta^{\star}>0$, the following condition must be satisfied:
\begin{equation}
\beta^{\star}=\sum_{i=1}^{N}\frac{A_{i}B_{i}+C_{i}D_{i}-\left(A_{i}^{2}+C_{i}^{2}\right)}{\sqrt{A_{i}^{2}+C_{i}^{2}}}>0.\label{eq:beta>0}
\end{equation}

2) $\alpha^{\star}\neq0$, $\beta^{\star}=0$ ; Since $\alpha^{\star}\neq0$,
we find $\lambda_{1}^{\star}=0$ from \ac{KKT} condition 5 in (\ref{eq:KKT conditions}).
The condition $\alpha^{\star}>0$ only holds if
\begin{equation}
\alpha^{\star}=\sum_{i=1}^{N}\frac{A_{i}B_{i}+C_{i}D_{i}-\left(B_{i}^{2}+D_{i}^{2}\right)}{\sqrt{B_{i}^{2}+D_{i}^{2}}}>0.\label{eq:alpha>0}
\end{equation}

3) $\alpha^{\star}=0$, $\beta^{\star}=0$ ; In this case, from \ac{KKT}
condition 7 in (\ref{eq:KKT conditions}) we deduce (\ref{eq:equation lambda})
shown on the next page.
\begin{figure*}[tbh]
\begin{align}
f(\lambda_{1}^{\star}) & \triangleq\sum_{i=1}^{N}\frac{\left(A_{i}-B_{i}\right)\left(\lambda_{1}^{\star}A_{i}+\left(1-\lambda_{1}^{\star}\right)B_{i}\right)+\left(C_{i}-D_{i}\right)\left(\lambda_{1}^{\star}C_{i}+\left(1-\lambda_{1}^{\star}\right)D_{i}\right)}{\sqrt{\left(\lambda_{1}^{\star}A_{i}+\left(1-\lambda_{1}^{\star}\right)B_{i}\right)^{2}+\left(\lambda_{1}^{\star}C_{i}+\left(1-\lambda_{1}^{\star}\right)D_{i}\right)^{2}}}=0.\label{eq:equation lambda}\\
\hline \nonumber 
\end{align}

\end{figure*}

Since $\left\{ A_{i}\right\} $, $\left\{ B_{i}\right\} $, $\left\{ C_{i}\right\} $
and $\left\{ D_{i}\right\} $ are independent \acp{RV} distributed
according to $\mathcal{N}\left(0,\frac{1}{2}\right)$, it can be shown
using the \ac{CLT} that it is extremely unlikely for conditions (\ref{eq:beta>0})
and (\ref{eq:alpha>0}) to be satisfied for large values of $N$.
Hence, solving the optimization problem (\ref{eq:STANDARD op problem})
reduces to solving the equation (\ref{eq:equation lambda}). However,
considering the fact that the function in (\ref{eq:Lagrange(1) function})
is convex in $\lambda_{1}$ and since $f(\lambda_{1}^{\star}=1)>0$
and $f(\lambda_{1}^{\star}=0)<0$, the solution to (\ref{eq:equation lambda})
must be unique. Since (\ref{eq:equation lambda}) does not admit an
analytical solution, it can be solved numerically to find $\lambda_{1}^{\star}$.

After the optimum $\lambda_{1}^{\star}$ has been determined, $\left\{ \nu_{i}^{\star}\right\} $
can be obtained via (\ref{eq:optimal nu}). Substituting $\nu_{i}^{\star}$
into (\ref{eq:optimum theta_R}) and (\ref{eq:optimum theta_I}),
we have 
\begin{equation}
\theta_{i}^{\mathcal{R}^{\star}}=\frac{\lambda_{1}^{\star}A_{i}+\left(1-\lambda_{1}^{\star}\right)B_{i}}{\sqrt{\left(\lambda_{1}^{\star}A_{i}+\left(1-\lambda_{1}^{\star}\right)B_{i}\right)^{2}+\left(\lambda_{1}^{\star}C_{i}+\left(1-\lambda_{1}^{\star}\right)D_{i}\right)^{2}}},\label{eq:optimum theta_R-1}
\end{equation}
for all $i=1,2,\dots,N$, and 
\begin{equation}
\theta_{i}^{\mathcal{I}^{\star}}=\frac{\lambda_{1}^{\star}C_{i}+\left(1-\lambda_{1}^{\star}\right)D_{i}}{\sqrt{\left(\lambda_{1}^{\star}A_{i}+\left(1-\lambda_{1}^{\star}\right)B_{i}\right)^{2}+\left(\lambda_{1}^{\star}C_{i}+\left(1-\lambda_{1}^{\star}\right)D_{i}\right)^{2}}},\label{eq:optimum theta_I-1}
\end{equation}
for all $i=1,2,\dots,N$.

The optimization procedure can be summarized as follows; first, parameters
$\left\{ A_{i}\right\} $, $\left\{ B_{i}\right\} $, $\left\{ C_{i}\right\} $
and $\left\{ D_{i}\right\} $ are given by 
\[
A_{i}=\pm h_{m,i}^{\mathcal{R}},\;B_{i}=\pm h_{n,i}^{\mathcal{I}},\;C_{i}=\mp h_{m,i}^{\mathcal{I}},\;D_{i}=\pm h_{n,i}^{\mathcal{R}},
\]
for all $i=1,2,\dots,N,$ where the polarity of the real part of the
desired received signal determines the sign used for defining parameters
$\left\{ A_{i}\right\} $ and $\left\{ C_{i}\right\} $, and the sign
used for $\left\{ B_{i}\right\} $ and $\left\{ D_{i}\right\} $ depends
on the polarity of the imaginary part of the received signal. Then,
$\lambda_{1}^{\star}$ is derived by solving the equation (\ref{eq:equation lambda}).
After this, the reflection coefficients $\left\{ \theta_{i}^{\star}\right\} $
can be calculated using (\ref{eq:optimum theta_R-1}) and (\ref{eq:optimum theta_I-1}).

\begin{figure}[t]
\begin{centering}
\includegraphics[scale=0.4]{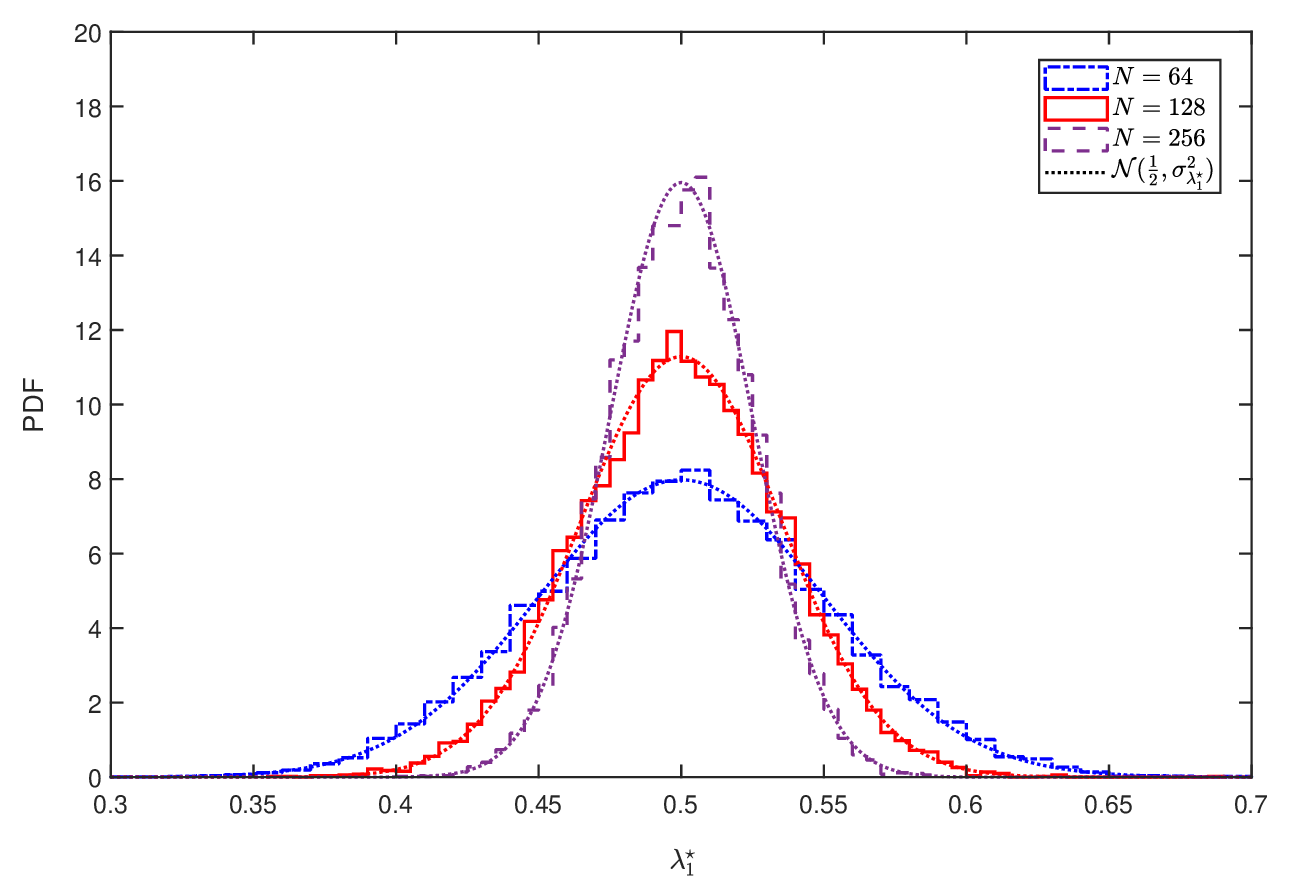}
\par\end{centering}
\caption{Histogram of the parameter $\lambda_{1}^{\star}$. \label{fig:Histogram-of-lambda1}}
\end{figure}

It is worth noting that there is a symmetry between $\lambda_{1}^{\star}$
and $\lambda_{2}^{\star}=1-\lambda_{1}^{\star}$ in (\ref{eq:equation lambda});
considering the fact that all variables $\left\{ A_{i}\right\} $,
$\left\{ B_{i}\right\} $, $\left\{ C_{i}\right\} $ and $\left\{ D_{i}\right\} $
are \acp{RV} identically distributed as $\mathcal{N}\left(0,\frac{1}{2}\right)$,
this reveals that $\lambda_{1}^{\star}$ and $\lambda_{2}^{\star}$
have equal mean, i.e., $\mathbb{E}\left\{ \lambda_{1}^{\star}\right\} =\mathbb{E}\left\{ \lambda_{2}^{\star}\right\} $,
which results in $\mathbb{E}\left\{ \lambda_{1}^{\star}\right\} =\frac{1}{2}$.
To gain further insights, we present the histogram of the optimum
$\lambda_{1}^{\star}$ in Fig.~\ref{fig:Histogram-of-lambda1}. This
figure shows the average number of occurrences of the value of $\lambda_{1}^{\star}$
in a specific interval in the domain $\left[0,1\right]$. Here, we
used $10^{4}$ channel realizations, and for each channel realization
(\ref{eq:equation lambda}) has been numerically solved to find the
optimum $\lambda_{1}^{\star}$. It can be observed that $\lambda_{1}^{\star}$
closely follows a Gaussian with mean $\frac{1}{2}$, i.e., $\lambda_{1}^{\star}\sim\mathcal{N}\left(\frac{1}{2},\sigma_{\lambda_{1}^{\star}}^{2}\right)$,
and that the variance $\sigma_{\lambda_{1}^{\star}}^{2}$ decreases
with an increasing number of \ac{RIS} elements $N$, such that it
can be neglected for large values of $N$ (e.g., with $N=256$, we
have $\sigma_{\lambda_{1}^{\star}}^{2}\approxeq6.2\times10^{-4}$).

\begin{figure}[t]
\begin{centering}
\includegraphics[scale=0.6]{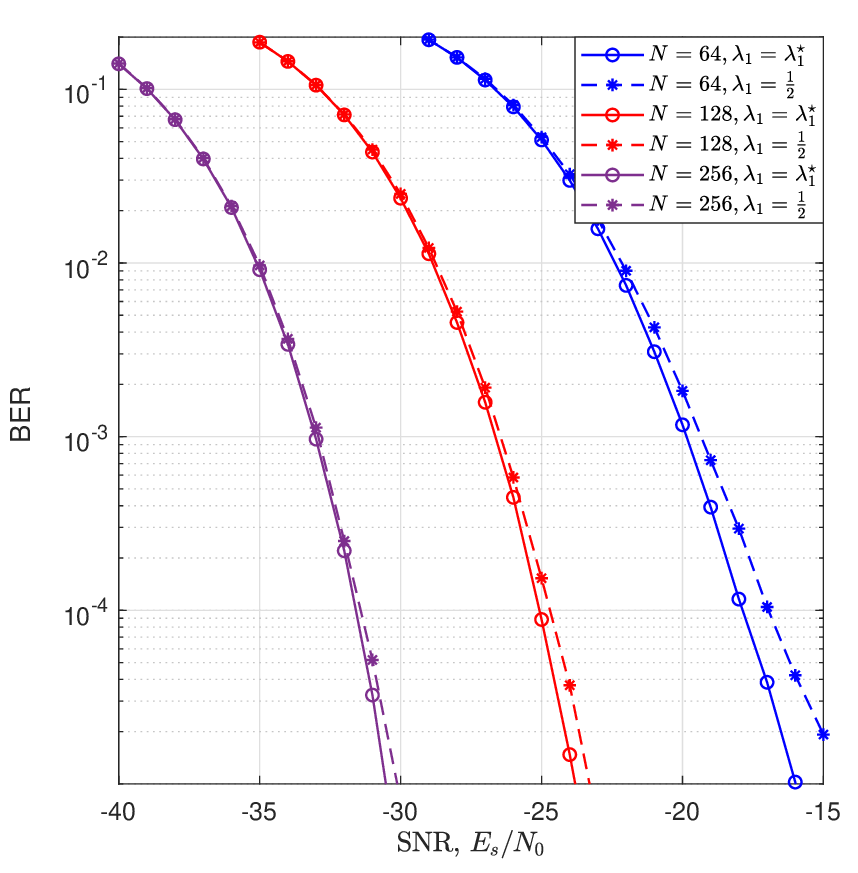}
\par\end{centering}
\caption{Impact of unoptimized $\lambda_{1}=\frac{1}{2}$ on the error rate
performance of the \ac{RIS-RQSSK} system without polarity bits, for
the case $N_{r}=4$.\label{fig:Impact-of-imperfect}}
\end{figure}

The previous observation will help us to derive an upper bound on
the error rate performance of the \ac{RIS-RQSSK} system with optimized
$\lambda_{1}^{\star}$. To see this, we present the \ac{BER} performance
of the \ac{RIS-RQSSK} system in Fig.~\ref{fig:Impact-of-imperfect}.
In this figure, for each value of $N$ we plot the simulation result
for the system with optimized $\lambda_{1}^{\star}$ and also for
a system which simply uses $\lambda_{1}=\frac{1}{2}$. As expected,
the performance of the system with $\lambda_{1}=\frac{1}{2}$ provides
an upper bound on the performance of the system with optimized $\lambda_{1}^{\star}$.
This upper bound becomes very tight at low \ac{SNR}, and also for
larger values of $N$. In practice, the optimized $\lambda_{1}^{\star}$
is exactly equal to $\frac{1}{2}$ only when $m=n$, i.e., the same
antenna is selected for both the real and imaginary parts.

\section{Performance Analysis\label{sec:Performance-Analysis}}

\subsection{RQSSK system without polarity bits\label{subsec:RQSSK-system-(without}}

In this section, we analyze the theoretical \ac{ABEP} of the \ac{RIS-RQSSK}
system. For ease of exposition, we will initially assume that there
are no polarity bits, i.e., $2\log_{2}N_{r}$ bits are transmitted
per \ac{IM} symbol, while the polarities are fixed and are not ``detected''
at the receiver. This analysis will focus on the \ac{GD} receiver
given by (\ref{eq:GD m_hat}) and (\ref{eq:GD n_hat}). Here we only
perform the analysis for the detection of the antenna $m$ with active
real part; due to the inherent symmetry in the expressions, it is
easy to show that the \ac{ABEP} expression for the detection of the
antenna $n$ with active imaginary part is identical. Considering
(\ref{eq:GD m_hat}), the \ac{PEP} associated with the selected antenna
$m$ and the detected antenna $\hat{m}\ne m$ is given by 
\begin{align}
 & \mbox{PEP}_{\mathrm{SSK}}\left(m,\hat{m}\right)=\Pr\left(\left(y_{m}^{\mathcal{R}}\right)^{2}<\left(y_{\hat{m}}^{\mathcal{R}}\right)^{2}\right)\nonumber \\
 & =\Pr\left\{ \left(\sqrt{E_{s}}\sum_{i=1}^{N}\left(h_{m,i}^{\mathcal{R}}\theta_{i}^{\mathcal{R}}-h_{m,i}^{\mathcal{I}}\theta_{i}^{\mathcal{I}}\right)+n_{m}^{\mathcal{R}}\right)^{2}<\right.\nonumber \\
 & \quad\left.\left(\sqrt{E_{s}}\sum_{i=1}^{N}\left(h_{\hat{m},i}^{\mathcal{R}}\theta_{i}^{\mathcal{R}}-h_{\hat{m},i}^{\mathcal{I}}\theta_{i}^{\mathcal{I}}\right)+n_{\hat{m}}^{\mathcal{R}}\right)^{2}\right\} .\label{eq:PEP}
\end{align}
Considering $\left\{ \theta_{i}^{\mathcal{R}}\right\} $ and $\left\{ \theta_{i}^{\mathcal{I}}\right\} $
given in (\ref{eq:optimum theta_R-1}) and (\ref{eq:optimum theta_I-1}),
and based on our discussion in the previous section regarding the
average value of $\lambda_{1}^{\star}$, the \ac{PEP} can be upper
bounded as 
\begin{align}
 & \mbox{PEP}_{\mathrm{SSK}}\left(m,\hat{m}\right)\leq\nonumber \\
 & \Pr\left(\left(\sqrt{E_{s}}Y+n_{m}^{\mathcal{R}}\right)^{2}<\left(\sqrt{E_{s}}\hat{Y}+n_{\hat{m}}^{\mathcal{R}}\right)^{2}\right),\label{eq:PEP Y and Y_hat}
\end{align}
where $Y=\sum_{i=1}^{N}Y_{i}=\sum_{i=1}^{N}\frac{A_{i}^{2}+C_{i}^{2}+A_{i}B_{i}+C_{i}D_{i}}{\sqrt{\left(A_{i}+B_{i}\right)^{2}+\left(C_{i}+D_{i}\right)^{2}}}$,
$\hat{Y}=\sum_{i=1}^{N}\hat{Y}_{i}=\sum_{i=1}^{N}\left(\hat{A}_{i}\theta_{i}^{\mathcal{R}}+\hat{C}_{i}\theta_{i}^{\mathcal{I}}\right)$,
and we define $\hat{A}_{i}=h_{\hat{m},i}^{\mathcal{R}}$ and $\hat{C}_{i}=-h_{\hat{m},i}^{\mathcal{I}}$
to simplify the notation. According to the \ac{CLT}, $Y$ and $\hat{Y}$
both follow a normal distribution, i.e., $Y\sim\mathcal{N}\left(N\mathbb{E}\left\{ Y_{i}\right\} ,N\mathbb{V}\left\{ Y_{i}\right\} \right)$
and $\hat{Y}\sim\mathcal{N}\left(N\mathbb{E}\left\{ \hat{Y}_{i}\right\} ,N\mathbb{V}\left\{ \hat{Y}_{i}\right\} \right)$.
Hence, the expected value and the variance of $Y_{i}$ and $\hat{Y}_{i}$
need to be evaluated. Since $B_{i}$ and $D_{i}$ depend on the selected
antenna $n$ with active imaginary part, we need to consider three
different cases: (i) $m\neq n$, $\hat{m}\neq n$, (ii) $m\neq n$,
$\hat{m}=n$, and (iii) $m=n$, $\hat{m}\neq n$. For all three cases,
the expected value and variance of $Y_{i}$ and $\hat{Y}_{i}$ can
be derived analytically. Hence, by introducing \acp{RV} $Z_{1}=\sqrt{E_{s}}Y+n_{m}^{\mathcal{R}}$
and $Z_{2}=\sqrt{E_{s}}\hat{Y}+n_{\hat{m}}^{\mathcal{R}}$, the following
theorems can be used to further analyze the \ac{PEP}.
\begin{thm}
For case (i) where we have $m\neq n$ and $\hat{m}\neq n$, $Z_{1}$
and $Z_{2}$ follow a normal distribution as \label{thm:case-(i)}
\[
Z_{1}\sim\mathcal{N}\left(\frac{N\sqrt{\pi E_{s}}}{2\sqrt{2}},\frac{\left(6-\pi\right)NE_{s}}{8}+\frac{N_{0}}{2}\right)
\]
and 
\[
Z_{2}\sim\mathcal{N}\left(0,\frac{NE_{s}}{2}+\frac{N_{0}}{2}\right).
\]
\end{thm}
\begin{IEEEproof}
The proof is provided in Appendix \ref{sec:proof of theorem case(i)}.
\end{IEEEproof}
\begin{thm}
For case (ii) where we have $m\neq n$ and $\hat{m}=n$, $Z_{1}$
and $Z_{2}$ follow a normal distribution as \label{thm:case-(ii)}
\[
Z_{1}\sim\mathcal{N}\left(\frac{N\sqrt{\pi E_{s}}}{2\sqrt{2}},\frac{\left(6-\pi\right)NE_{s}}{8}+\frac{N_{0}}{2}\right)
\]
and 
\[
Z_{2}\sim\mathcal{N}\left(0,\frac{NE_{s}}{4}+\frac{N_{0}}{2}\right).
\]
\end{thm}
\begin{IEEEproof}
See Appendix \ref{sec:proof of theorem case(ii)}.
\end{IEEEproof}

\begin{thm}
For case (iii) where we have $m=n$ and $\hat{m}\neq n$, $Z_{1}$
and $Z_{2}$ follow a normal distribution as \label{thm:case-(iii)}
\[
Z_{1}\sim\mathcal{N}\left(\frac{N\sqrt{\pi E_{s}}}{2\sqrt{2}},\frac{\left(4-\pi\right)NE_{s}}{8}+\frac{N_{0}}{2}\right)
\]
and 
\[
Z_{2}\sim\mathcal{N}\left(0,\frac{NE_{s}}{2}+\frac{N_{0}}{2}\right).
\]
\end{thm}
\begin{IEEEproof}
See Appendix \ref{sec:proof of theorem case(iii)}.\footnote{From Theorems 1-3, we see that the signal at the selected antenna,
$Z_{1}$, and the signal at a non-selected antenna, $Z_{2}$, follow
normal distributions as $Z_{1}\sim\mathcal{N}\left(\mu,\sigma_{1}^{2}\right)$
and $Z_{2}\sim\left(0,\sigma_{2}^{2}\right)$, where in a noise-free
system (i.e., $\mathrm{SNR}\rightarrow\infty$), we have $\mu,\sigma_{1}^{2},\sigma_{2}^{2}\propto N$.
We know that for a normal RV $Z\sim\left(\mu,\sigma^{2}\right)$,
we have $\frac{1}{\sqrt{2\pi\sigma^{2}}}\int_{\mu-n\sigma}^{\mu+n\sigma}f_{Z}\left(z\right)\mathrm{d}z=1-\epsilon$,
where $\epsilon<<1$ for $n\geq2$, e.g., with $n=5$ we obtain $\epsilon\approxeq5.7\times10^{-7}$.
Considering this fact and since $\frac{\mu}{\sigma_{1}},\frac{\mu}{\sigma_{2}}\propto\sqrt{N}>>1$,
then it follows that it is highly unlikely to erroneously detect the
SSK symbol, i.e., maximizing the signal at the target antenna provides
a near-optimal approach.}
\end{IEEEproof}
Therefore, the \ac{PEP} can be expressed as 
\begin{align}
\mbox{PEP}_{\mathrm{SSK}}\left(m,\hat{m}\right)\leq & \frac{N_{r}-2}{N_{r}}\Pr\left(Q<0|m\neq n,\hat{m}\neq n\right)\nonumber \\
 & +\frac{1}{N_{r}}\Pr\left(Q<0|m\neq n,\hat{m}=n\right)\nonumber \\
 & +\frac{1}{N_{r}}\Pr\left(Q<0|m=n,\hat{m}\neq n\right),\label{eq:PEP full}
\end{align}
where $Q=Q_{1}-Q_{2}$ is the difference of the two non-central chi-square
\acp{RV} $Q_{1}=Z_{1}^{2}$ and $Q_{2}=Z_{2}^{2}$ each having one
degree of freedom. It can be easily proved that $Z_{1}$ and $Z_{2}$
are uncorrelated, i.e., $\mathbb{E}\left\{ Z_{1}Z_{2}\right\} =\mathbb{E}\left\{ Z_{1}\right\} \mathbb{E}\left\{ Z_{2}\right\} =0$,
and are therefore independent (recall that $Z_{1}$ and $Z_{2}$ are
both normal \acp{RV}). The \ac{PDF} and \ac{CDF} of $Q$ can both
be expressed in the form of an infinite series expansion; however,
for odd values of the number of degrees of freedom the resulting expression
is prohibitively complex \cite{simon2002probability}. Hence, we implement
numerical methods and approximations to obtain the exact or approximated
upper bounds on the \ac{PEP} results. The Gil-Pelaez inversion formula
\cite[Eq. 4.4.1]{mathai1992quadratic} is a numerical method that
can be used to calculate the \ac{CDF} of $Q$ as 
\begin{equation}
F_{Q}\left(q\right)=\Pr\left(Q<q\right)=\frac{1}{2}-\frac{1}{\pi}\int_{0}^{\infty}t^{-1}\left(\phi_{Q}\left(t\right)e^{-jtq}\right)^{\mathcal{I}}\mathrm{d}t,\label{eq:Gil-Paelz}
\end{equation}
where $\phi_{Q}\left(t\right)$ is the \ac{CF} of $Q$ which can
be derived from the \ac{LT} of the \ac{PDF} (see (\ref{eq:general LT})
in Appendix \ref{sec:proof of theorem case(i)}, i.e., $\phi_{Q}\left(t\right)=\mathcal{L}_{-jt}\left(f_{Q}\left(q\right)\right)$.
Hence, the exact upper bound expression on the \ac{PEP} in (\ref{eq:PEP full})
can be obtained by numerically evaluating the integral in (\ref{eq:Gil-Paelz})
for $q=0$. However, to obtain further insight into the resulting
\ac{PEP}, we use Pearson's approximation approach \cite{imhof1961computing},
where the distribution of a linear combination of non-central quadratic
form of standard normal \acp{RV} $\left\{ X_{i}\right\} $, i.e.
$Q'=\sum_{i=1}^{n}\delta_{i}\left(X_{i}+b_{i}\right)^{2}$, is approximated
by that of a central chi-square \ac{RV}, i.e.,
\begin{equation}
Q'\approx\frac{c_{3}}{c_{2}}\chi_{v}^{2}-\frac{c_{2}^{2}}{c_{3}}+c_{1},\label{eq:approx_chi}
\end{equation}
where $\approx$ here means ``approximately distributed'',
\begin{equation}
c_{k}=\sum_{i=1}^{n}\delta_{i}^{k}\left(1+kb_{i}^{2}\right),\ k=1,2,3,\label{eq:c_k}
\end{equation}
and 
\begin{equation}
v=\frac{c_{2}^{3}}{c_{3}^{2}}.\label{eq:DoF}
\end{equation}
We use this approach as it is simple yet remarkably accurate in both
tails of the distribution. Hence, we obtain 
\begin{equation}
\Pr\left(Q'<q\right)\approxeq\Pr\left(\chi_{v}^{2}<\bar{q}\right),\label{eq:Pr_Q'}
\end{equation}
where 
\[
\bar{q}=\left(q-c_{1}\right)\sqrt{\frac{v}{c_{2}}}+v.
\]
Expressing $Q$ as the quadratic form $Q'$, for case (i) where we
have $m\neq n$ and $\hat{m}\neq n$, we obtain 
\[
\delta_{1}=\frac{\left(6-\pi\right)NE_{s}}{8}+\frac{N_{0}}{2},\ \delta_{2}=-\frac{NE_{s}}{2}-\frac{N_{0}}{2},
\]
\[
b_{1}^{2}=\frac{\pi N^{2}E_{s}}{\left(6-\pi\right)NE_{s}+4N_{0}},\ b_{2}=0.
\]
These expressions can be similarly derived for cases (ii) and (iii).
Substituting the corresponding variables for each case into (\ref{eq:c_k})
and (\ref{eq:DoF}), the details of the approximated chi-square \ac{RV}
can be derived accordingly. Hence, considering $q=0$ in (\ref{eq:Pr_Q'}),
then for each case we have 
\begin{align}
 & \Pr\left(Q<0\right)\approxeq\Pr\left(\chi_{v}^{2}<\bar{q}\right)\nonumber \\
 & =F\left(\bar{q};v\right)=\frac{\gamma\left(\frac{v}{2},\frac{\bar{q}}{2}\right)}{\Gamma\left(\frac{v}{2}\right)}=\frac{\gamma\left(\frac{v}{2},\frac{-c_{1}\sqrt{\nicefrac{v}{c_{2}}}+v}{2}\right)}{\Gamma\left(\frac{v}{2}\right)},\label{eq:PEP_approx}
\end{align}
where $F\left(x;k\right)$ is the \ac{CDF} of a chi-square \ac{RV}
with $k$ degrees of freedom, and $\gamma\left(s,x\right)=\int_{0}^{x}u^{s-1}e^{-s}du$
is the lower incomplete gamma function. Calculating (\ref{eq:PEP_approx})
for all three cases in (\ref{eq:PEP full}), we obtain an approximate
upper bound on the \ac{PEP}.

Considering the \ac{SNR} range $N\frac{E_{s}}{N_{0}}\ll1$ (which
is of interest for large values of $N$; note that $N\gg1$), then
we have 
\begin{align*}
 & v\approxeq\frac{2\left(\pi N^{2}\frac{E_{s}}{N_{0}}+4\right)^{3}}{9\left(\pi N^{2}\frac{E_{s}}{N_{0}}\right)^{2}},\\
 & \bar{q}\approxeq\frac{2\left(\pi N^{2}\frac{E_{s}}{N_{0}}+4\right)^{3}}{9\left(\pi N^{2}\frac{E_{s}}{N_{0}}\right)^{2}}-\frac{1}{6}\left(\pi N^{2}\frac{E_{s}}{N_{0}}+4\right).
\end{align*}
Note that these approximate values of $v$ and $\bar{q}$ are equal
in all three cases (i)-(iii). Then, for \ac{SNR} values in the range
$N^{2}\frac{E_{s}}{N_{0}}\gg1$ (which is also valid in the range
of interest for large values of $N$), we can write 
\[
v\approxeq\frac{2\pi N^{2}}{9}\frac{E_{s}}{N_{0}},\ \bar{q}\approxeq\frac{1}{4}v.
\]
Then, it can be shown that the Chernoff bound on the lower tail of
the \ac{CDF} of a chi-square distribution $F\left(x;k\right)$, where
$x=zk$, is given by
\[
F\left(zk;k\right)\leq\left(ze^{1-z}\right)^{\frac{k}{2}},
\]
 hence, (\ref{eq:PEP_approx}) can be expressed as 
\begin{equation}
\mbox{PEP}_{\mathrm{SSK}}\left(m\rightarrow\hat{m}\right)\leq\Pr\left(Q<0\right)\lessapprox\left(\frac{2}{e^{\nicefrac{3}{8}}}\right)^{\frac{-2\pi N^{2}}{9}\frac{E_{s}}{N_{0}}},\label{eq:upper bound on PEP}
\end{equation}
which indicates the nature of the error rate performance enhancement
that is obtained by increasing $N$. It is worth noting that (\ref{eq:upper bound on PEP})
is similar in form to the Chernoff bound on the Q-function. Hence,
we can conclude that the RIS-RQSSK system model behaves like a Gaussian
channel with an average power gain that is proportional to $N^{2}$. 

Finally, according to the union bound, the \ac{ABEP} can be expressed
as 
\[
\mbox{ABEP}\leq\frac{1}{N_{r}\log_{2}N_{r}}\sum_{m}\sum_{\hat{m}}\mbox{PEP}_{\mathrm{SSK}}\left(m\rightarrow\hat{m}\right)e\left(m\rightarrow\hat{m}\right),
\]
where $e\left(m\rightarrow\hat{m}\right)$ is the Hamming distance
between the binary representations of symbols $m$ and $\hat{m}$.
It is worth pointing out that the \ac{PEP} is independent of $m$
and $\hat{m}$; hence the \ac{ABEP} can be re-expressed as 
\begin{equation}
\mbox{ABEP}\leq\frac{N_{r}}{2}\mbox{PEP}_{\mathrm{SSK}}.\label{eq:ABEP}
\end{equation}

\noindent 
\begin{figure}[tbh]
\begin{centering}
\includegraphics[scale=0.6]{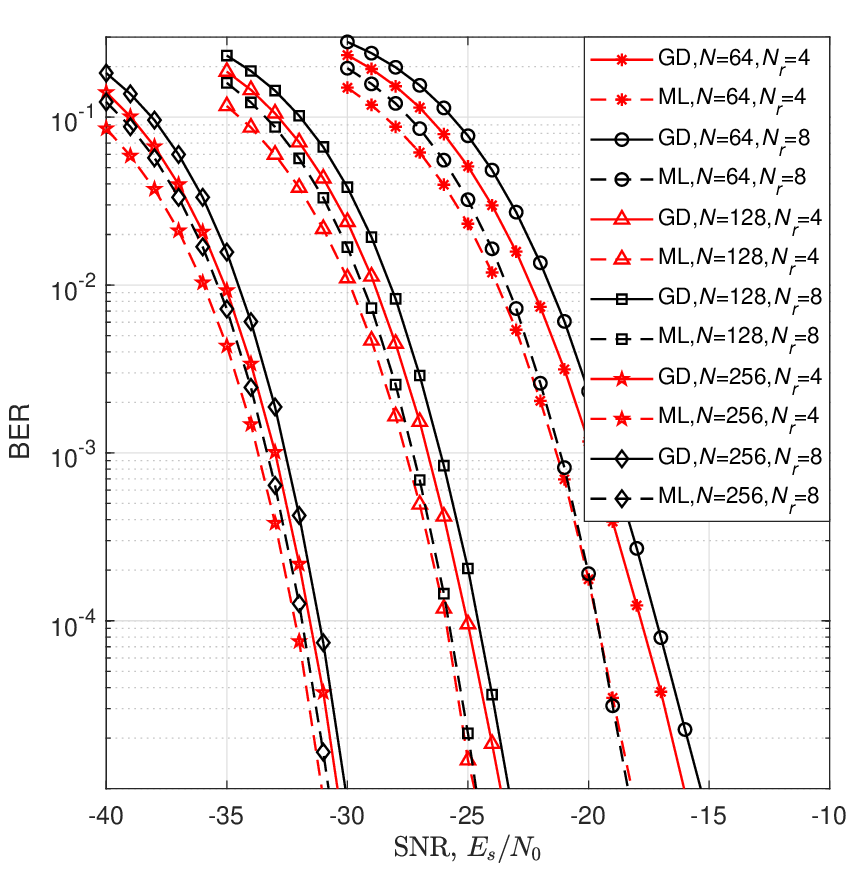}
\par\end{centering}
\caption{Comparison of the performance of the GD and ML detector in the proposed
RIS-RQSSK system.\label{fig:GD_ML_compare}}
\end{figure}

\subsection{RQSSK including polarity bits\label{subsec:RQSSK-including-polarity}}

Next, we consider the case where two bits in each data packet of $2\left(\log_{2}N_{r}+1\right)$
bits is transmitted by the polarities of the real and imaginary ``active''
received signals. Similar to the case without polarity bits, we focus
only on the real part (the analysis for the imaginary part is the
same by symmetry). The pair $\left(m,d_{m}\right)$ represents the
super-symbol comprised of \ac{SSK} bits and the polarity bit, where
$m$ denotes the antenna index with active real part and $d_{m}$
denotes the polarity bit. Hence, considering erroneous and correct
detection of the antenna index separately, an upper bound on the \ac{ABEP}
can be derived as 
\begin{align}
\mbox{ABEP}\leq & \frac{1}{2N_{r}\left(\log_{2}N_{r}+1\right)}\nonumber \\
 & \sum_{(m,d)}\sum_{(\hat{m},\hat{d})}\mbox{PEP}\left(m,d\rightarrow\hat{m},\hat{d}_{m}\right)e\left(m,d\rightarrow\hat{m},\hat{d}_{m}\right)\nonumber \\
= & \frac{1}{\log_{2}N_{r}+1}\left[1-\left(N_{r}-1\right)\mbox{PEP}_{\mathrm{SSK}}\right]\mbox{PEP}_{\mathrm{POL}|m=\hat{m}}\nonumber \\
 & +\frac{N_{r}\log_{2}N_{r}}{2\left(\log_{2}N_{r}+1\right)}\mbox{PEP}_{\mathrm{SSK}}\left(1-\mbox{PEP}_{\mathrm{POL}|m\neq\hat{m}}\right)\nonumber \\
 & +\frac{\frac{N_{r}}{2}\log_{2}N_{r}+N_{r}-1}{\log_{2}N_{r}+1}\mbox{PEP}_{\mathrm{SSK}}\mbox{PEP}_{\mathrm{POL}|m\neq\hat{m}},\label{eq:ABEP total}
\end{align}
where $\mbox{PEP}_{\mathrm{POL}|m=\hat{m}}$ and $\mbox{PEP}_{\mathrm{POL}|m\neq\hat{m}}$
are the average \ac{PEP} associated with the pair of polarity symbols
$\left(d_{m},\hat{d}_{m}\right)$ conditioned on correct and erroneous
detection of $m$, respectively, which are given by (\ref{eq:ave. PEP_d1})
and (\ref{eq:ave. PEP_d2}) shown on the next page.
\begin{figure*}[tbh]
\begin{align}
\mbox{PEP}_{\mathrm{POL}|m=\hat{m}} & =\frac{1}{N_{r}}\mathbb{E}_{Y}\left\{ \mbox{Q}\left(\sqrt{Y^{2}\frac{2E_{s}}{N_{0}}}\right)|m=n\right\} +\frac{N_{r}-1}{N_{r}}\mathbb{E}_{Y}\left\{ \mbox{Q}\left(\sqrt{Y^{2}\frac{2E_{s}}{N_{0}}}\right)|m\neq n\right\} ,\label{eq:ave. PEP_d1}\\
\mbox{PEP}_{\mathrm{POL}|m\neq\hat{m}} & =\frac{1}{N_{r}-1}\mathbb{E}_{\hat{Y}}\left\{ \mbox{Q}\left(\sqrt{\hat{Y}^{2}\frac{2E_{s}}{N_{0}}}\right)|\hat{m}=n\right\} +\frac{N_{r}-2}{N_{r}-1}\mathbb{E}_{\hat{Y}}\left\{ \mbox{Q}\left(\sqrt{\hat{Y}^{2}\frac{2E_{s}}{N_{0}}}\right)|\hat{m}\neq n\right\} .\label{eq:ave. PEP_d2}\\
\hline \nonumber 
\end{align}
\end{figure*}
 To calculate these average \acp{PEP}, we use Craig's alternative
formula for the Q-function \cite{craig1991new}. Hence, for an RV
$V$ we can write
\begin{equation}
\mathbb{E}_{V}\left\{ \mbox{Q}\left(\sqrt{V}\right)\right\} =\frac{1}{\pi}\int_{0}^{\nicefrac{\pi}{2}}\mathcal{M}_{V}\left(\frac{-1}{2\sin^{2}\phi}\right)\mathrm{d}\phi,\label{eq:ave. Q}
\end{equation}
where $\mathcal{M}_{V}\left(s\right)$ is the \ac{MGF} of \ac{RV}
$V$. For each case in (\ref{eq:ave. PEP_d1}) and (\ref{eq:ave. PEP_d2}),
the \ac{MGF} of the received \ac{SNR} can be calculated from the
\ac{LT} of the distribution function (see (\ref{eq:general LT})
in Appendix \ref{sec:proof of theorem case(i)}). Finally, by computing
the integral in (\ref{eq:ave. Q}) for all Q-functions in (\ref{eq:ave. PEP_d1})
and (\ref{eq:ave. PEP_d2}) and substituting (\ref{eq:ave. PEP_d1})
and (\ref{eq:ave. PEP_d2}), and exact (based on Gil-Pelaez formula)
or approximate (based on Pearson approach or Chernoff bound) upper
bound on $\mbox{PEP}_{\mathrm{SSK}}$ into (\ref{eq:ABEP total}),
the desired \ac{ABEP} for the \ac{RIS-RQSSK} system with polarity
bits can be derived.

\section{Numerical Results\label{sec:Numerical-Results}}

\noindent In this section, we demonstrate the performance of the proposed
\ac{RIS-RQSSK} system through numerical results. As benchmarks, we
consider the most prominent recently proposed \ac{RIS}-based schemes
which incorporate the concept of \ac{SM}, namely \ac{RIS-RQRM} \cite{yuan2021receive}
and \ac{RIS-SM} \cite{basar2020reconfigurable} systems. First, in
Fig.~\ref{fig:GD_ML_compare} we compare the performance of the RIS-RQSSK
system where the GD and ML detector are implemented at the receiver.
It can be seen that the GD performs fairly close to the ML detector
in terms of the error rate. The performance gap between GD and ML
detector reduces with an increasing number of RIS elements.

\begin{figure*}[tbh]
\begin{centering}
\includegraphics[scale=0.44]{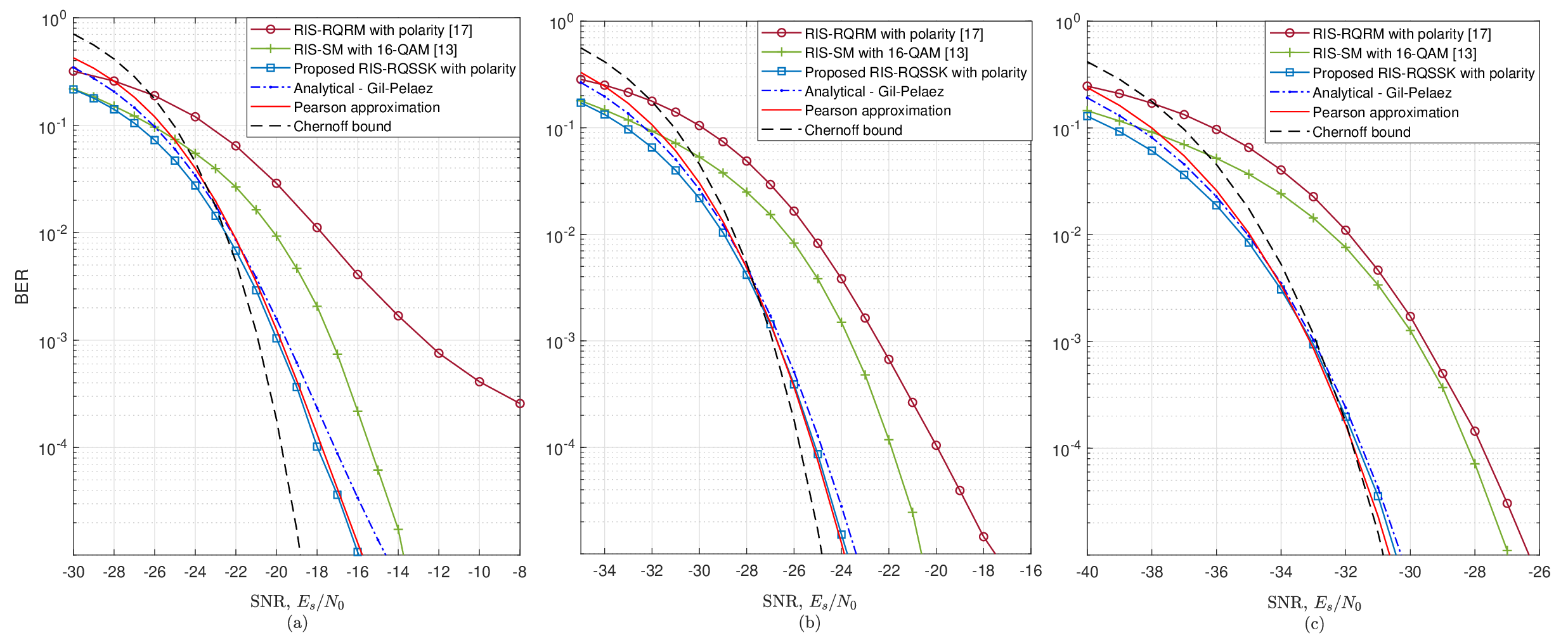}
\par\end{centering}
\caption{Analytical and simulation results of the proposed \ac{RIS-RQSSK}
system, and comparison of the performance with that of \acs{RIS-RQRM}
and \acs{RIS-SM} systems. Here $N_{r}=4$ and $R=6$. (a) $N=64$,
(b) $N=128$, (c) $N=256$.\label{fig:Analytical-and-numerical-Nr4}}
\end{figure*}

Then, Fig.~\ref{fig:Analytical-and-numerical-Nr4} shows the \ac{BER}
performance of the proposed \ac{RIS-RQSSK} system as well as that
of the benchmark schemes with $N_{r}=4$, for different values of
the number of \ac{RIS} elements $N$. The phase shifts for the proposed
system are optimized according to the solution provided in Section~\ref{sec:Problem-Formulation}.
Both the \ac{RIS-RQSSK} and \ac{RIS-RQRM} systems exploit the polarity
of the signals at the receiver to transmit two additional bits; however,
in order to have a fair comparison, the \ac{RIS-SM} system uses 16-QAM
modulation in order to compensate for the additional bits transmitted
by the quadrature branch. Hence, the data rate is $R=6$~\ac{bpcu}
in all of the considered schemes. We perform \ac{GD} at the receiver
in all scenarios to detect the spatial symbols, and hence \ac{CSI}
is not required in the \ac{RIS-RQSSK} and \ac{RIS-RQRM} systems,
while \ac{RIS-SM} needs to know the channel amplitudes in order to
detect the QAM symbols, i.e., partial channel knowledge is required.
 It can be observed that the proposed system significantly outperforms
the \ac{RIS-RQRM} system which is an alternative quadrature-based
approach for spatial modulation. However, in \ac{RIS-RQRM}, the \ac{RIS}
elements are divided into two equal groups each of which targets either
the real or the imaginary part of the selected receive antenna, which
clearly deteriorates the performance of the system. The gain of the
proposed scheme over \ac{RIS-RQRM} is 6.2~dB and 4.1~dB for systems
with $N=128$ and $N=256$, respectively, at a \ac{BER} of $10^{-5}$.
In the system with $N=64$, a larger gain is obtained, since in the
\ac{RIS-RQRM} system, the number of \ac{RIS} elements assisting
each activated real/imaginary part is $N=32$, which results in an
error floor in the moderate \ac{BER} range. The proposed \ac{RIS-RQSSK}
also provides superior performance to \ac{RIS-SM}. This superiority
is due to the fact that the \ac{RIS-SM} system uses 16-QAM modulation
to achieve $R=6$~\ac{bpcu}, which results in a diminishing \ac{BER}
performance due to the smaller minimum Euclidean distance between
different constellation points. The proposed \ac{RIS-RQSSK} system
achieves 2.4~dB, 3.2~dB and 3.5~dB performance improvement over
the \ac{RIS-SM} system for systems with $N=64$, $N=128$ and $N=256$,
respectively. In this figure, we also plot the analytical and approximate
results discussed in the previous section. The exact results based
on the Gil-Pelaez inversion formula validate the simulation curves
for the proposed system. The tiny difference between the simulation
and Gil-Pelaez curves is caused by the fact that we use $\lambda_{1}=\frac{1}{2}$
in the theoretical analysis, while the optimized $\lambda_{1}^{\star}$
is used in the simulations. This difference becomes smaller with increasing
$N$, as expected (c.f., Fig.~\ref{fig:Impact-of-imperfect}). In
addition, it can be seen that the results based on the Pearson approximation
approach are very close to that of the Gil-Pelaez formula, especially
for larger values of $N$. However, as expected, the resulting Chernoff
bounds are not quite valid at small values of $N$, as in this case
the conditions on the \ac{SNR} range are not satisfied. However,
for large values of $N$, i.e., $N\geq256$, the gap between the Chernoff
bound and the theoretical curve becomes small enough that it can be
neglected.

\begin{figure*}[tbh]
\begin{centering}
\includegraphics[scale=0.44]{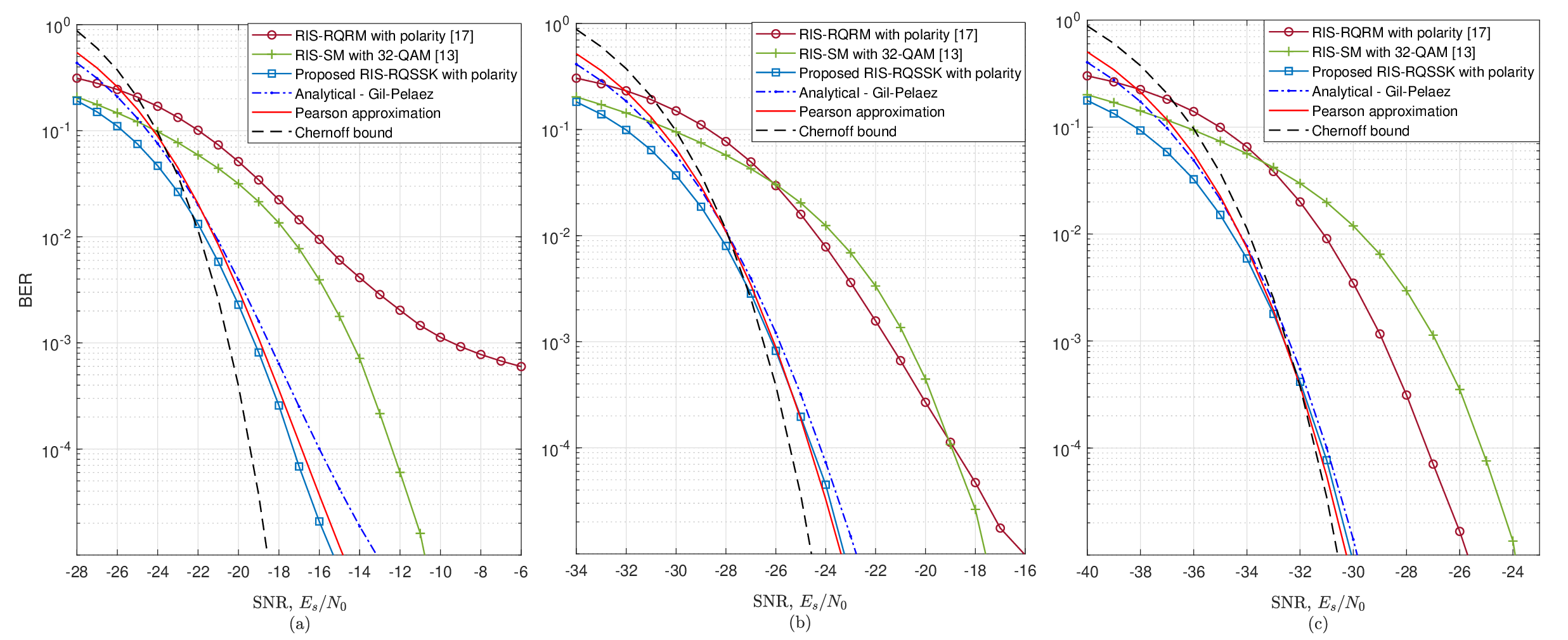}
\par\end{centering}
\caption{Analytical and simulation results of the proposed \ac{RIS-RQSSK}
system, and comparison of the performance with that of \ac{RIS-RQRM}
and \ac{RIS-SM} systems. Here $N_{r}=8$ and $R=8$. (a) $N=64$,
(b) $N=128$, (c) $N=256$.\label{fig:Analytical-and-simulation-Nr8}}
\end{figure*}

In Fig.~\ref{fig:Analytical-and-simulation-Nr8}, we plot the corresponding
results for a system with $N_{r}=8$. Here the \ac{RIS-SM} system
uses 32-QAM, and hence the data rate is $R=8$~\ac{bpcu} in all scenarios.
The results show that the \ac{BER} of the proposed \ac{RIS-RQSSK}
system significantly outperforms that of the other schemes. Also,
we see that this improvement increases with increasing $N_{r}$\footnote{It is worth pointing out that the LTE-Advanced standard supports eight
antennas in the downlink and four antennas in the uplink \cite{etsiTR36913},
hence we used four and eight receive antennas in our simulations,
even though the gains from our proposed scheme become even higher
when the number of receive antennas is increased beyond eight.}. This is due to the fact that the number of \ac{RIS} elements in
the \ac{RIS-RQRM} system becomes smaller compared to the number of
receive antennas, and therefore the \ac{RIS} system cannot efficiently
perform beamforming toward the target receive antenna. Besides, the
use of 32-QAM in the \ac{RIS-SM} system results in a very small minimum
Euclidean distance which degrades the performance of the system, such
that it performs even worse than \ac{RIS-RQRM} in the case where
$N=256$. The proposed \ac{RIS-RQSSK} designs provide 7.2~dB and
4.4~dB improvement over \ac{RIS-RQRM} with $N=128$ and $N=256$,
respectively, and 4.5~dB, 5.6~dB and 6.2~dB gain over \ac{RIS-SM}
with $N=64$, $N=128$ and $N=256$, respectively. It is worth noting
that the \ac{RIS-SM} scheme requires partial \ac{CSI} at the receiver,
while the \ac{GD} in the proposed \ac{RIS-RQSSK} system performs
\ac{CSI}-free detection. Finally, it can be seen that compared to
the system with $N_{r}=4$, the gap between the theoretical and numerical
curves is slightly increased. The reason for this is that in a system
with $N_{r}=4$, the coincidence of $m=n$ is more likely to occur,
which, as previously mentioned, yields $\lambda_{1}^{\star}=\frac{1}{2}$,
the same value that is used in the theoretical analysis.

\section{Conclusion\label{sec:Conclusion}}

In this paper, we introduced a novel \ac{IM} scheme for \ac{RIS}-assisted
wireless communications, called \ac{RIS-RQSSK}, in which \ac{SSK}
was performed independently in both the real and imaginary dimensions.
The key advantage of this approach is that \emph{all} \ac{RIS} elements
perform beamforming onto the real and imaginary part of the received
signal at the selected antennas, respectively. Therefore, in addition
to realizing an enhancement in the spectral efficiency, the error
rate performance is also improved. We also defined and formulated
a max-min optimization problem to maximize the instantaneous \ac{SNR}
components at the selected antennas. We provided an analytical solution
for this non-convex problem, such that the multi-variable optimization
problem can be transformed to a simple single-variable equation. We
analyzed the \ac{ABEP} of the proposed scheme, and derived analytical
upper bounds on the approximate \ac{ABEP}. The \ac{BER} performance
of the proposed \ac{RIS-RQSSK} system has been demonstrated through
extensive numerical simulations. Our numerical results have shown
that the proposed \ac{RIS-RQSSK} system enormously outperforms the
recent prominent benchmark schemes such as \ac{RIS-SM} and \ac{RIS-RQRM},
thus providing an approach for \ac{RIS}-aided wireless communications
which exhibits a high energy efficiency without compromising on spectral
efficiency. This low-powered single-antenna AP system can be a potential
candidate for small cells in cellular networks. Moreover, in addition
to the use of a simple GD at the receiver, reliable communications
can be achieved with simpler channel codes which require lower complexity
at the receiver. As a result, the proposed approach can be adopted
to serve user devices in the downlink that are energy-constrained
and require low-complexity receiver algorithms. Finally, an interesting
direction of future research is to develop the proposed system in
order to support IQ modulation as well as index modulation to obtain
higher data rates.

\appendices{}

\section{Proof of Theorem \ref{thm:case-(i)}\label{sec:proof of theorem case(i)}}

Here we analyze the expected value and variance of $Y_{i}$ and $\hat{Y}_{i}$
where $m\neq n$ and $\hat{m}\neq n$ (case (i)), which are further
used to evaluate the expected value and variance of $Z_{1}$ and $Z_{2}$.

\subsection{Expected value of $Y_{i}=\frac{A_{i}^{2}+C_{i}^{2}+A_{i}B_{i}+C_{i}D_{i}}{\sqrt{\left(A_{i}+B_{i}\right)^{2}+\left(C_{i}+D_{i}\right)^{2}}}$}

$Y_{i}$ consists of four terms defined as 
\[
W_{1}\triangleq\frac{A^{2}}{\sqrt{Z}},\;W_{2}\triangleq\frac{C^{2}}{\sqrt{Z}},\;W_{3}\triangleq\frac{AB}{\sqrt{Z}},\;W_{4}\triangleq\frac{CD}{\sqrt{Z}},
\]
where we define $Z\triangleq(A+B)^{2}+(C+D)^{2}$ and also we omit
the index $i$ in order to simplify the notation. In the following,
we evaluate the expected value of each term individually.

According to the law of total expectation, the expected value of $W_{1}$
can be given by 
\begin{align}
 & \mathbb{E}\left\{ W_{1}\right\} =\mathbb{E}_{A}\left\{ \mathbb{E}_{W_{1}|A}\left\{ W_{1}|A\right\} \right\} \nonumber \\
 & =\mathbb{E}_{A}\left\{ \mathbb{E}_{W_{1}|A}\left\{ \frac{A^{2}}{\sqrt{Z}}|A\right\} \right\} =\mathbb{E}_{A}\left\{ A^{2}\mathbb{E}_{Z|A}\left\{ Z^{-\frac{1}{2}}|A\right\} \right\} ,\label{eq:E=00007BW1=00007Ddef}
\end{align}
where $\mathbb{E}_{Z|A}\left\{ Z^{-\frac{1}{2}}|A\right\} $ is the
inverse-fractional moment of $Z$ where $A$ is given, i.e., where
$A$ is a constant. For a given $A$, the \ac{RV} $(A+B)$ is normal
with mean $A$ and variance $\frac{1}{2}$, i.e., $\left(A+B\right)\sim\mathcal{N}\left(A,\frac{1}{2}\right)$,
and also $\left(C+D\right)$ is distributed according to $\mathcal{N}\left(0,1\right)$.
Hence, the \ac{RV} $(Z|A)$ is the sum of two independent chi-square
\acp{RV} each having one degree of freedom, i.e.,
\[
\left(Z|A\right)\sim\chi_{1}^{2}\left(A^{2}\right)+\chi_{1}^{2}.
\]
To compute the inverse and fractional moment of $\left(Z|A\right)$,
we use the definition of the gamma function to write \cite{mathai1992quadratic}
\begin{align}
\mathbb{E}_{Z|A}\left\{ Z^{-c}|A\right\}  & =\mathbb{E}_{Z|A}\left\{ \frac{1}{\Gamma\left(c\right)}\int_{0}^{\infty}s^{c-1}e^{-sZ}\mathrm{d}s|A\right\} \nonumber \\
 & =\frac{1}{\Gamma\left(c\right)}\int_{0}^{\infty}s^{c-1}\mathbb{E}_{Z|A}\left\{ e^{-sZ}|A\right\} \mathrm{d}s.\label{eq:fractional moment formula}
\end{align}
However, $\mathbb{E}_{Z|A}\left\{ e^{-sZ}|A\right\} $ is the \ac{MGF}
of $\left(Z|A\right)$, $\mathcal{M}_{Z|A}\left(-s\right)$, or the
\ac{LT} of $f_{Z}\left(Z|A\right)$, $\mathcal{L}_{s}\left(f_{Z}\left(Z|A\right)\right)$.
We know that the \ac{LT} of the \ac{PDF} of the sum of independent
\acp{RV} is the multiplication of the \ac{LT} of their individual
\acp{PDF}, and that the \ac{LT} of the \ac{PDF} of an \ac{RV}
$X=\sum_{i=1}^{n}X_{i}^{2}$ with $X_{i}\sim\mathcal{N}\left(\mu_{i},\sigma^{2}\right)$
is given by
\begin{equation}
\mathcal{L}_{s}\left(f_{X}(X)\right)=\left(\frac{1}{1+2\sigma^{2}s}\right)^{\frac{n}{2}}\exp\left(\frac{-\mu^{2}s}{1+2\sigma^{2}s}\right),\label{eq:general LT}
\end{equation}
where $\mu^{2}=\sum_{i=1}^{n}\mu_{i}^{2}$. Hence, the \ac{LT} of
$f_{Z}\left(Z|A\right)$ can be expressed as 
\begin{equation}
\mathcal{L}_{s}\left(f_{Z}\left(Z|A\right)\right)=\left(\frac{1}{1+2s}\right)^{\frac{1}{2}}\left(\frac{1}{1+s}\right)^{\frac{1}{2}}\exp\left(\frac{-A^{2}s}{1+s}\right).\label{eq:Laplas(Z|A)}
\end{equation}
Then, (\ref{eq:E=00007BW1=00007Ddef}) can be written as (\ref{eq:E=00007BW1=00007D})
shown on the next page.
\begin{figure*}[tbh]
\begin{align}
\mathbb{E}\left\{ W_{1}\right\}  & =\mathbb{E}_{A}\left\{ \frac{A^{2}}{\Gamma\left(\frac{1}{2}\right)}\int_{0}^{\infty}s^{\frac{1}{2}-1}\left(\frac{1}{1+2s}\right)^{\frac{1}{2}}\left(\frac{1}{1+s}\right)^{\frac{1}{2}}\exp\left(\frac{-A^{2}s}{1+s}\right)\mathrm{d}s\right\} \nonumber \\
 & =\frac{1}{\Gamma\left(\frac{1}{2}\right)}\int_{-\infty}^{\infty}A^{2}\left(\int_{0}^{\infty}s^{\frac{1}{2}-1}\left(\frac{1}{1+2s}\right)^{\frac{1}{2}}\left(\frac{1}{1+s}\right)^{\frac{1}{2}}\exp\left(\frac{-A^{2}s}{1+s}\right)\mathrm{d}s\right)f_{A}\left(A\right)\mathrm{d}A.\label{eq:E=00007BW1=00007D}\\
\hline \nonumber 
\end{align}

\end{figure*}
 Considering $f_{A}\left(A\right)=\frac{1}{\Gamma\left(\frac{1}{2}\right)}\exp\left(-A^{2}\right)$,
and changing the order of integrals we have 
\begin{align*}
\mathbb{E}\left\{ W_{1}\right\} = & \frac{1}{\Gamma^{2}\left(\frac{1}{2}\right)}\int_{0}^{\infty}s^{\frac{1}{2}-1}\left(\frac{1}{1+2s}\right)^{\frac{1}{2}}\left(\frac{1}{1+s}\right)^{\frac{1}{2}}\\
 & \left(\int_{-\infty}^{\infty}A^{2}\exp\left(-A^{2}\frac{1+2s}{1+s}\right)\mathrm{d}A\right)\mathrm{d}s.
\end{align*}
Since $\int_{-\infty}^{\infty}x^{2}\exp\left(-\frac{x^{2}}{2\sigma^{2}}\right)\mathrm{d}x=\Gamma\left(\frac{1}{2}\right)\sigma^{2}\sqrt{2\sigma^{2}}$,
we have $\int_{-\infty}^{\infty}A^{2}\exp\left(-A^{2}\frac{1+2s}{1+s}\right)\mathrm{d}A=\frac{\Gamma\left(\frac{1}{2}\right)}{2}\left(\frac{1+s}{1+2s}\right)^{\frac{3}{2}}$;
hence 
\begin{align*}
\mathbb{E}\left\{ W_{1}\right\} = & \frac{1}{2\Gamma\left(\frac{1}{2}\right)}\int_{0}^{\infty}s^{\frac{1}{2}-1}\frac{1+s}{\left(1+2s\right)^{2}}\mathrm{d}s\\
= & \frac{1}{2\Gamma\left(\frac{1}{2}\right)}\left(\frac{1}{2}\int_{0}^{\infty}s^{\frac{1}{2}-1}\left(1+2s\right)^{-1}\mathrm{d}s\right.\\
 & \left.+\frac{1}{2}\int_{0}^{\infty}s^{\frac{1}{2}-1}\left(1+2s\right)^{-2}\mathrm{d}s\right).
\end{align*}
Considering the type-2 beta function defined as $B\left(\alpha,\beta\right)=\int_{0}^{\infty}\frac{t^{\alpha-1}}{\left(1+t\right)^{\alpha+\beta}}\mathrm{d}t=\frac{\Gamma\left(\alpha\right)\Gamma\left(\beta\right)}{\Gamma\left(\alpha+\beta\right)}$,
with some minor manipulation we obtain 
\begin{align*}
\mathbb{E}\left\{ W_{1}\right\}  & =\frac{1}{2\Gamma\left(\frac{1}{2}\right)}\left(\frac{1}{2\sqrt{2}}\frac{\Gamma\left(\frac{1}{2}\right)\Gamma\left(\frac{1}{2}\right)}{\Gamma\left(1\right)}+\frac{1}{2\sqrt{2}}\frac{\Gamma\left(\frac{1}{2}\right)\Gamma\left(\frac{3}{2}\right)}{\Gamma\left(2\right)}\right)\\
 & =\frac{3\sqrt{\pi}}{8\sqrt{2}}.
\end{align*}
By symmetry it is clear that $\mathbb{E}\left\{ W_{2}\right\} =\mathbb{E}\left\{ W_{1}\right\} $.

Next we determine $\mathbb{E}\left\{ W_{3}=\frac{AB}{\sqrt{Z}}\right\} $.
Considering the law of total expectation and expanding this for multiple
\acp{RV}, we have 
\begin{align*}
\mathbb{E}\left\{ W_{3}\right\}  & =\mathbb{E}_{(A,B)}\left\{ \mathbb{E}_{W_{3}|(A,B)}\left\{ W_{3}|(A,B)\right\} \right\} \\
 & =\mathbb{E}_{A}\left\{ A\mathbb{E}_{B}\left\{ B\mathbb{E}_{Z|(A,B)}\left\{ Z^{-\frac{1}{2}}|(A,B)\right\} \right\} \right\} .
\end{align*}
Now, given $(A,B)$, $Z$ is the sum of the central chi-square \ac{RV}
$\left(C+D\right)^{2}$ and the constant $\left(A+B\right)^{2}$.
Hence, $Z=\bar{Z}+\left(A+B\right)^{2}$, where $\bar{Z}=(C+D)^{2}$,
then we have $f_{Z}\left(Z|(A,B)\right)=f_{\bar{Z}}\left(Z-\left(A+B\right)^{2}\right)$.
Therefore, 
\begin{align*}
\mathcal{L}_{s}\left(f_{Z}\left(Z|(A,B)\right)\right) & =\mathcal{L}_{s}\left(f_{\bar{Z}}\left(Z-\left(A+B\right)^{2}\right)\right)\\
 & =\exp\left(-\left(A+B\right)^{2}s\right)\mathcal{L}_{s}\left(f_{\bar{Z}}\left(\bar{Z}\right)\right).
\end{align*}
Hence, given (\ref{eq:general LT}), the \ac{LT} of $f_{Z}\left(Z|(A,B)\right)$
can be expressed as
\begin{align}
\mathcal{L}_{s}\left(f_{Z}\left(Z|(A,B)\right)\right) & =\exp\left(-\left(A+B\right)^{2}s\right)\mathcal{L}_{s}\left(f_{\bar{Z}}\left(\bar{Z}\right)\right)\nonumber \\
 & =\left(\frac{1}{1+2s}\right)^{\frac{1}{2}}\exp\left(-\left(A+B\right)^{2}s\right).\label{eq:Laplace(Z|A,B)}
\end{align}
Using (\ref{eq:fractional moment formula}), we have 
\begin{align*}
\mathbb{E}\left\{ W_{3}\right\} = & \mathbb{E}_{A}\left\{ A\mathbb{E}_{B}\left\{ \frac{B}{\Gamma\left(\frac{1}{2}\right)}\int_{0}^{\infty}s^{\frac{1}{2}-1}\left(\frac{1}{1+2s}\right)^{\frac{1}{2}}\right.\right.\\
 & \exp\left(-\left(A+B\right)^{2}s\right)\mathrm{d}s\Biggr\}\Biggr\}.
\end{align*}
Using $f_{A}\left(A\right)=\frac{1}{\Gamma\left(\frac{1}{2}\right)}\exp\left(-A^{2}\right)$
and $f_{B}\left(B\right)=\frac{1}{\Gamma\left(\frac{1}{2}\right)}\exp\left(-B^{2}\right)$,
and after some manipulation we can write {\small{}
\begin{align}
\mathbb{E}\left\{ W_{3}\right\} = & \frac{1}{\Gamma^{3}\left(\frac{1}{2}\right)}\int_{0}^{\infty}s^{\frac{1}{2}-1}\left(\frac{1}{1+2s}\right)^{\frac{1}{2}}\Biggl[\int_{-\infty}^{\infty}A\exp\Biggl(-A^{2}\frac{1+2s}{1+s}\Biggr)\nonumber \\
 & \Biggl(\int_{-\infty}^{\infty}B\exp\Biggl(-\frac{\left(B+\frac{As}{1+s}\right)^{2}}{\frac{1}{1+s}}\Biggr)\mathrm{d}B\Biggr)\mathrm{d}A\Biggr]\mathrm{d}s.\label{eq:E=00007BW3=00007D}
\end{align}
}Since $\int_{-\infty}^{\infty}x\exp\left(-\frac{\left(x-\mu\right)^{2}}{2\sigma^{2}}\right)\mathrm{d}x=\Gamma\left(\frac{1}{2}\right)\mu\sqrt{2\sigma^{2}}$,
the inner integral over $B$ can be evaluated as 
\[
\int_{-\infty}^{\infty}B\exp\left(-\frac{\left(B+\frac{As}{1+s}\right)^{2}}{\frac{1}{1+s}}\right)\mathrm{d}B=-\Gamma\left(\frac{1}{2}\right)\frac{As}{1+s}\left(\frac{1}{1+s}\right)^{\frac{1}{2}}.
\]
Substituting this into (\ref{eq:E=00007BW3=00007D}), the expected
value of $W_{3}$ is given by 
\begin{align*}
\mathbb{E}\left\{ W_{3}\right\} = & \frac{-1}{\Gamma^{2}\left(\frac{1}{2}\right)}\int_{0}^{\infty}s^{\frac{3}{2}-1}\left(\frac{1}{1+2s}\right)^{\frac{1}{2}}\left(\frac{1}{1+s}\right)^{\frac{3}{2}}\\
 & \left(\int_{-\infty}^{\infty}A^{2}\exp\left(-A^{2}\frac{1+2s}{1+s}\right)\mathrm{d}A\right)\mathrm{d}s\\
= & \frac{-1}{2\Gamma\left(\frac{1}{2}\right)}\int_{0}^{\infty}s^{\frac{3}{2}-1}\left(\frac{1}{1+2s}\right)^{2}\mathrm{d}s\\
= & \frac{-1}{4\sqrt{2}\Gamma\left(\frac{1}{2}\right)}B\left(\frac{3}{2},\frac{1}{2}\right)=-\frac{\sqrt{\pi}}{8\sqrt{2}}.
\end{align*}
Also, by symmetry we have $\mathbb{E}\{W_{4}\}=\mathbb{E}\{W_{3}\}$.
Therefore, the expected value of $Y_{i}$ is given by 
\begin{equation}
\mathbb{E}\left\{ Y_{i}\right\} =2\left(\mathbb{E}\left\{ W_{1}\right\} +\mathbb{E}\left\{ W_{3}\right\} \right)=\frac{\sqrt{\pi}}{2\sqrt{2}}.\label{eq:E=00007BY_i=00007D}
\end{equation}

\subsection{Variance of $Y_{i}=\frac{A_{i}^{2}+C_{i}^{2}+A_{i}B_{i}+C_{i}D_{i}}{\sqrt{\left(A_{i}+B_{i}\right)^{2}+\left(C_{i}+D_{i}\right)^{2}}}$}

The variance of $Y_{i}$ is given by 
\[
\mathbb{V}\left\{ Y_{i}\right\} =\mathbb{E}\left\{ Y_{i}^{2}\right\} -\mathbb{E}\left\{ Y_{i}\right\} ^{2}.
\]
However, $\mathbb{E}\left\{ Y_{i}\right\} $ is given in (\ref{eq:E=00007BY_i=00007D});
hence, only the expected value of $Y_{i}^{2}$ needs to be evaluated.
We separate the expression of $Y_{i}^{2}$ into 4 terms as 
\[
U_{1}\triangleq\frac{A^{4}+A^{2}B^{2}+2A^{3}B}{Z},\;U_{2}\triangleq\frac{C^{4}+C^{2}D^{2}+2C^{3}D}{Z},
\]
\[
U_{3}\triangleq\frac{2A^{2}CD+2A^{2}C^{2}}{Z},\;U_{4}\triangleq\frac{2ABC^{2}+2ABCD}{Z}.
\]
We can analyze the expected values of $U_{1}$ to $U_{4}$ separately.
However, it is trivial that $\mathbb{E}\left\{ U_{1}\right\} =\mathbb{E}\left\{ U_{2}\right\} $.
Therefore, in the following we calculate the expected values of $U_{1}$,
$U_{3}$ and $U_{4}$.
\begin{itemize}
\item Expected value of $U_{1}=\frac{A^{4}+A^{2}B^{2}+2A^{3}B}{Z}$
\end{itemize}
According to the law of total expectation, we have 
\begin{align*}
 & \mathbb{E}\left\{ U_{1}\right\} =\\
 & \mathbb{E}_{(A,B)}\left\{ \left(A^{4}+A^{2}B^{2}+2A^{3}B\right)\mathbb{E}_{Z|(A,B)}\left\{ Z^{-1}|(A,B)\right\} \right\} .
\end{align*}
From (\ref{eq:fractional moment formula}) and (\ref{eq:Laplace(Z|A,B)}),
$\mathbb{E}\left\{ U_{1}\right\} $ can be expressed as 
\begin{align*}
 & \mathbb{E}\left\{ U_{1}\right\} =\\
 & \frac{1}{\Gamma^{2}\left(\frac{1}{2}\right)}\int_{0}^{\infty}\left(\frac{1}{1+2s}\right)^{\frac{1}{2}}\Biggl[\int_{-\infty}^{\infty}\int_{-\infty}^{\infty}\left(A^{4}+A^{2}B^{2}+2A^{3}B\right)\\
 & \exp\Biggl(-\frac{\left(B+\frac{As}{1+s}\right)^{2}}{\frac{1}{1+s}}\Biggr)\exp\biggl(-A^{2}\frac{1+2s}{1+s}\biggr)\mathrm{d}B\mathrm{d}A\Biggr]\mathrm{d}s\\
 & =\frac{\Gamma\left(\frac{5}{2}\right)}{\Gamma\left(\frac{1}{2}\right)}\int_{0}^{\infty}\frac{1}{\left(1+2s\right)^{3}}\mathrm{d}s+\frac{1}{4}\int_{0}^{\infty}\frac{1}{\left(1+2s\right)^{2}}\mathrm{d}s\\
 & =\frac{3}{4}\cdot\frac{1}{4}+\frac{1}{4}\cdot\frac{1}{2}=\frac{5}{16}.
\end{align*}

\begin{itemize}
\item Expected value of $U_{3}=\frac{2A^{2}CD+2A^{2}C^{2}}{Z}$
\end{itemize}
$\mathbb{E}\left\{ U_{3}\right\} $ can be evaluated as (\ref{eq:E(U3)})
shown on the next page.
\begin{figure*}[tbh]
\begin{align}
\mathbb{E}\left\{ U_{3}\right\} = & \mathbb{E}_{(A,C,D)}\left\{ \left(2A^{2}CD+2A^{2}C^{2}\right)\mathbb{E}_{Z|(A,C,D)}\left\{ Z^{-1}|(A,C,D)\right\} \right\} \nonumber \\
= & \frac{2}{\Gamma^{3}\left(\frac{1}{2}\right)}\int_{-\infty}^{\infty}\int_{-\infty}^{\infty}\int_{-\infty}^{\infty}\left(A^{2}CD+A^{2}C^{2}\right)\left(\int_{0}^{\infty}\mathcal{L}_{s}\left(f_{Z|(A,C,D)}\right)\mathrm{d}s\right)\exp\left(-D^{2}\right)\exp\left(-C^{2}\right)\exp\left(-A^{2}\right)\mathrm{d}D\mathrm{d}C\mathrm{d}A\nonumber \\
= & \frac{2}{\Gamma^{3}\left(\frac{1}{2}\right)}\int_{0}^{\infty}\left(\frac{1}{1+s}\right)^{\frac{1}{2}}\left(\int_{-\infty}^{\infty}A^{2}\exp\left(-A^{2}\frac{1+2s}{1+s}\right)\mathrm{d}A\right)\nonumber \\
 & \left(\int_{-\infty}^{\infty}\int_{-\infty}^{\infty}\left(CD+C^{2}\right)\exp\left(-\frac{\left(D+\frac{Cs}{1+s}\right)^{2}}{\frac{1}{1+s}}\right)\exp\left(-C^{2}\frac{1+2s}{1+s}\right)\mathrm{d}D\mathrm{d}C\right)\mathrm{d}s\nonumber \\
= & \frac{1}{2}\int_{0}^{\infty}\frac{1+s}{\left(1+2s\right)^{3}}\mathrm{d}s=\frac{3}{16}.\label{eq:E(U3)}\\
\hline \nonumber 
\end{align}

\end{figure*}

\begin{itemize}
\item Expected value of $U_{4}=\frac{2ABC^{2}+2ABCD}{Z}$
\end{itemize}
We can write 
\begin{align*}
 & \mathbb{E}\left\{ U_{4}\right\} =\\
 & \mathbb{E}_{(A,B,C)}\left\{ 2ABC\mathbb{E}_{(U_{4}/2ABC)|(A,B,C)}\left\{ \frac{C+D}{Z}|(A,B,C)\right\} \right\} .
\end{align*}
In contrast to the previous calculations where an inverse-fractional
moment of a quadratic function of an \ac{RV} or \acp{RV} were required,
here the first moment of a quotient of two functions of an \ac{RV}
needs to be evaluated. Positive integer moments of $R=\frac{Q_{1}}{Q_{2}}$
can be calculated by \cite[Eq. 4.5a.2]{mathai1992quadratic}
\begin{equation}
\mathbb{E}\left\{ R^{c}\right\} =\frac{1}{\Gamma\left(c\right)}\int_{0}^{\infty}s_{2}^{c-1}\left.\left[\frac{\partial^{c}}{\partial s_{1}^{c}}\mathcal{M}_{Q_{1},Q_{2}}\left(s_{1},-s_{2}\right)\right]\right|_{s_{1}=0}\mathrm{d}s_{2},\label{eq:moments of ratio}
\end{equation}
where $\mathcal{M}_{Q_{1},Q_{2}}\left(s_{1},s_{2}\right)$ is the
joint \ac{MGF} of $Q_{1}$ and $Q_{2}$. Further, the joint \ac{MGF}
of quadratic forms $Q_{1}=\mathbf{x}^{T}\mathbf{A}_{1}\mathbf{x}+\mathbf{a}_{1}^{T}\mathbf{x}+d_{1}$
and $Q_{2}=\mathbf{x}^{T}\mathbf{A}_{2}\mathbf{x}+\mathbf{a}_{2}^{T}\mathbf{x}+d_{2}$,
where $\mathbf{x}\sim\mathcal{N}_{p}\left(\boldsymbol{\mu},\boldsymbol{\Sigma}\right)$
is a length-$p$ normal random vector, is given by (\ref{eq:joint MGF of Q1-Q2})
shown on the next page \cite[Eq. 3.2c.5]{mathai1992quadratic}.
\begin{figure*}[tbh]

\begin{align}
\mathcal{M}_{Q_{1},Q_{2}}\left(s_{1},s_{2}\right)= & \det\left(\mathbf{I}-2s_{1}\mathbf{A}_{1}\boldsymbol{\Sigma}-2s_{2}\mathbf{A}_{2}\boldsymbol{\Sigma}\right)^{-\frac{1}{2}}\exp\left[-\frac{1}{2}\left(\boldsymbol{\mu}^{T}\boldsymbol{\Sigma}^{-1}\boldsymbol{\mu}-2s_{1}d_{1}-2s_{2}d_{2}\right)+\right.\nonumber \\
 & \left.\frac{1}{2}\left(s_{1}\boldsymbol{\Sigma}\mathbf{a}_{1}+s_{2}\boldsymbol{\Sigma}\mathbf{a}_{2}+\boldsymbol{\mu}\right)^{T}\left(\mathbf{I}-2s_{1}\mathbf{A}_{1}\boldsymbol{\Sigma}-2s_{2}\mathbf{A}_{2}\boldsymbol{\Sigma}\right)^{-1}\boldsymbol{\Sigma}^{-1}\left(s_{1}\boldsymbol{\Sigma}\mathbf{a}_{1}+s_{2}\boldsymbol{\Sigma}\mathbf{a}_{2}+\boldsymbol{\mu}\right)\right].\label{eq:joint MGF of Q1-Q2}\\
\hline \nonumber 
\end{align}
\end{figure*}
 Hence, the joint \ac{MGF} of $Q_{1}=C+D$ and $Q_{2}=\left(A+B\right)^{2}+\left(C+D\right)^{2}$,
where $(A,B,C)$ is given, can be expressed as 
\begin{align*}
 & \mathcal{M}_{(Q_{1},Q_{2})|(A,B,C)}\left(s_{1},s_{2}\right)=\\
 & \frac{1}{\left(1-s_{2}\right)^{\frac{1}{2}}}\exp\left(-C^{2}+s_{2}\left(A+B\right)^{2}+\frac{\left(\frac{1}{2}s_{1}+C\right)^{2}}{\left(1-s_{2}\right)}\right).
\end{align*}
Substituting this into (\ref{eq:moments of ratio}) and considering
$c=1$, we can write 
\begin{align*}
 & \mathbb{E}_{(U_{4}/2ABC)|(A,B,C)}\left\{ \frac{C+D}{Z}|(A,B,C)\right\} =\\
 & \int_{0}^{\infty}\frac{C}{\left(1+s_{2}\right)^{\frac{3}{2}}}\exp\left(-C^{2}\frac{s_{2}}{1+s_{2}}\right)\exp\left(-\left(A+B\right)^{2}s_{2}\right)\mathrm{d}s_{2}.
\end{align*}
Then, $\mathbb{E}\left\{ U_{4}\right\} $ can be evaluated as (\ref{eq:E(U4)})
shown on the next page.
\begin{figure*}[tbh]

\begin{align}
\mathbb{E}\left\{ U_{4}\right\} = & \frac{2}{\Gamma^{3}\left(\frac{1}{2}\right)}\int_{-\infty}^{\infty}\int_{-\infty}^{\infty}\int_{-\infty}^{\infty}ABC\Biggl[\int_{0}^{\infty}\frac{C}{\left(1+s_{2}\right)^{\frac{3}{2}}}\exp\biggl(-C^{2}\frac{s_{2}}{1+s_{2}}\biggr)\exp\left(-\left(A+B\right)^{2}s_{2}\right)\mathrm{d}s_{2}\Biggr]\nonumber \\
 & \exp\left(-C^{2}\right)\exp\left(-B^{2}\right)\exp\left(-A^{2}\right)\mathrm{d}C\mathrm{d}B\mathrm{d}A\nonumber \\
= & \frac{2}{\Gamma^{3}\left(\frac{1}{2}\right)}\int_{0}^{\infty}\frac{1}{\left(1+s_{2}\right)^{\frac{3}{2}}}\Biggl[\int_{-\infty}^{\infty}\int_{-\infty}^{\infty}AB\exp\Biggl(-\frac{\Bigl(B+\frac{As_{2}}{1+s_{2}}\Bigr)^{2}}{\frac{1}{1+s_{2}}}\Biggr)\exp\left(-A^{2}\frac{1+2s_{2}}{1+s_{2}}\right)\mathrm{d}B\mathrm{d}A\Biggr]\nonumber \\
 & \biggl(\int_{-\infty}^{\infty}C^{2}\exp\Bigl(-C^{2}\frac{1+2s_{2}}{1+s_{2}}\Bigr)\mathrm{d}C\biggr)\mathrm{d}s_{2}\nonumber \\
= & -\frac{1}{2}\int_{0}^{\infty}\frac{s_{2}}{\left(1+2s_{2}\right)^{3}}\mathrm{d}s_{2}=-\frac{1}{16}.\label{eq:E(U4)}\\
\hline \nonumber 
\end{align}
\end{figure*}

Therefore, the variance of $Y_{i}$ is calculated as 
\begin{align*}
\mathbb{V}\left\{ Y_{i}\right\}  & =\mathbb{E}\left\{ Y_{i}^{2}\right\} -\mathbb{E}\left\{ Y_{i}\right\} ^{2}\\
 & =\left(2\cdot\frac{5}{16}+\frac{3}{16}-\frac{1}{16}\right)-\left(\frac{\sqrt{\pi}}{2\sqrt{2}}\right)^{2}=\frac{6-\pi}{8}.
\end{align*}

\subsection{Expected value of $\hat{Y}_{i}=\hat{A}_{i}\theta_{i}^{\mathcal{R}}+\hat{C}_{i}\theta_{i}^{\mathcal{I}}$}

Since $\theta_{i}^{\mathcal{R}}$ and $\theta_{i}^{\mathcal{I}}$
are independent of $\hat{A}_{i}$ and $\hat{C_{i}}$, then $\mathbb{E}\left\{ \hat{Y}_{i}\right\} $
is calculated as 
\[
\mathbb{E}\left\{ \hat{Y}_{i}\right\} =\mathbb{E}\left\{ \hat{A}_{i}\right\} \mathbb{E}\left\{ \theta_{i}^{\mathcal{R}}\right\} +\mathbb{E}\left\{ \hat{C}_{i}\right\} \mathbb{E}\left\{ \theta_{i}^{\mathcal{I}}\right\} =0.
\]

\subsection{Variance of $\hat{Y}_{i}=\hat{A}_{i}\theta_{i}^{\mathcal{R}}+\hat{C}_{i}\theta_{i}^{\mathcal{I}}$}

The variance of $\hat{Y}_{i}$ is given by 
\begin{align*}
\mathbb{V}\left\{ \hat{Y}_{i}\right\} = & \mathbb{E}\left\{ \hat{Y}_{i}^{2}\right\} \\
= & \mathbb{E}\left\{ \hat{A}_{i}^{2}\right\} \mathbb{E}\left\{ \left(\theta_{i}^{\mathcal{R}}\right)^{2}\right\} +\mathbb{E}\left\{ \hat{C}_{i}^{2}\right\} \mathbb{E}\left\{ \left(\theta_{i}^{\mathcal{I}}\right)^{2}\right\} \\
 & +2\mathbb{E}\left\{ \hat{A}_{i}\right\} \mathbb{E}\left\{ \hat{C}_{i}\right\} \mathbb{E}\left\{ \theta_{i}^{\mathcal{R}}\theta_{i}^{\mathcal{I}}\right\} \\
= & \frac{1}{2}\mathbb{E}\left\{ \left(\theta_{i}^{\mathcal{R}}\right)^{2}+\left(\theta_{i}^{\mathcal{I}}\right)^{2}\right\} =\frac{1}{2}.
\end{align*}

\subsection{Expected value and variance of $Z_{1}$ and $Z_{2}$}

Finally, the expected value and variance of $Z_{1}=\sqrt{E_{s}}Y+n_{m}^{\mathcal{R}}$
and $Z_{2}=\sqrt{E_{s}}\hat{Y}+n_{\hat{m}}^{\mathcal{R}}$ can be
derived as 
\begin{align*}
 & \mathbb{E}\left\{ Z_{1}\right\} =\mathbb{E}\left\{ \sqrt{E_{s}}Y+n_{m}^{\mathcal{R}}\right\} =\frac{N\sqrt{\pi E_{s}}}{2\sqrt{2}},\\
 & \mathbb{V}\left\{ Z_{1}\right\} =\mathbb{V}\left\{ \sqrt{E_{s}}Y+n_{m}^{\mathcal{R}}\right\} =\frac{\left(6-\pi\right)NE_{s}}{8}+\frac{N_{0}}{2},\\
 & \mathbb{E}\left\{ Z_{2}\right\} =\mathbb{E}\left\{ \sqrt{E_{s}}\hat{Y}+n_{m}^{\mathcal{R}}\right\} =0,\\
 & \mathbb{V}\left\{ Z_{2}\right\} =\mathbb{V}\left\{ \sqrt{E_{s}}\hat{Y}+n_{\hat{m}}^{\mathcal{R}}\right\} =\frac{NE_{s}}{2}+\frac{N_{0}}{2},
\end{align*}
thus completing the proof of Theorem \ref{thm:case-(i)}.

\section{Proof of Theorem \ref{thm:case-(ii)}\label{sec:proof of theorem case(ii)}}

Here we investigate the expected value and variance of $Y_{i}$ and
$\hat{Y}_{i}$ where $m\neq n$ and $\hat{m}=n$ (case (ii)).

In this case, the expected value and variance of $Y_{i}$ are equal
to the corresponding values derived for case (i); however, since $\hat{m}=n$,
then we have $\hat{A}_{i}=D_{i}$ and $\hat{C}_{i}=-B_{i}$. Hence,
$\hat{Y}_{i}$ is given by 
\[
\hat{Y}_{i}=\frac{D_{i}\left(A_{i}+B_{i}\right)-B_{i}\left(C_{i}+D_{i}\right)}{\sqrt{\left(A_{i}+B_{i}\right)^{2}+\left(C_{i}+D_{i}\right)^{2}}}.
\]
Therefore, $\mathbb{E}\left\{ \hat{Y}_{i}\right\} $ and $\mathbb{V}\left\{ \hat{Y}_{i}\right\} $
need to be evaluated. It is trivial that $\mathbb{E}\left\{ \hat{Y}_{i}\right\} =0$,
hence, in the following, we calculate $\mathbb{V}\left\{ \hat{Y}_{i}\right\} =\mathbb{E}\left\{ \hat{Y}_{i}^{2}\right\} $.

Considering the terms in $\hat{Y}_{i}^{2}$ (omitting index $i$),
i.e.,
\[
\hat{U}_{1}\triangleq\frac{D^{2}\left(A+B\right)^{2}}{Z},\ \hat{U}_{2}\triangleq\frac{B^{2}\left(C+D\right)^{2}}{Z},
\]
\[
\hat{U}_{3}\triangleq-\frac{2BD\left(A+B\right)\left(C+D\right)}{Z},
\]
expected value of $\hat{U}_{1}$ can be evaluated as (\ref{eq:E(Uhat1)})
shown on the next page,
\begin{figure*}[tbh]

\begin{align}
\mathbb{E}\left\{ \hat{U}_{1}\right\} = & \mathbb{E}_{(A+B,D)}\left\{ D^{2}\left(A+B\right)^{2}\mathbb{E}_{Z|(A+B,D)}\left\{ Z^{-1}|(A+B,D)\right\} \right\} \nonumber \\
= & \frac{1}{\sqrt{2}\Gamma^{2}\left(\frac{1}{2}\right)}\int_{-\infty}^{\infty}\int_{-\infty}^{\infty}D^{2}E^{2}\left(\int_{0}^{\infty}\mathcal{L}_{s}\left(f_{Z|(A+B,D)}\right)\mathrm{d}s\right)\exp\left(-D^{2}\right)\exp\left(-\frac{E^{2}}{2}\right)\mathrm{d}D\mathrm{d}E\nonumber \\
= & \frac{1}{\sqrt{2}\Gamma^{2}\left(\frac{1}{2}\right)}\int_{0}^{\infty}\left(\frac{1}{1+s}\right)^{\frac{1}{2}}\left(\int_{-\infty}^{\infty}D^{2}\exp\left(-D^{2}\frac{1+2s}{1+s}\right)\mathrm{d}D\right)\left(\int_{-\infty}^{\infty}E^{2}\exp\left(-E^{2}\frac{1+2s}{2}\right)\mathrm{d}E\right)\mathrm{d}s\nonumber \\
= & \frac{1}{2}\int_{0}^{\infty}\frac{1+s}{\left(1+2s\right)^{3}}\mathrm{d}s=\frac{3}{16}.\label{eq:E(Uhat1)}\\
\hline \nonumber 
\end{align}
\end{figure*}
 where we define $E=A+B$, which is distributed according to $\mathcal{N}\left(0,1\right)$.
By a similar calculation we can show that $\mathbb{E}\left\{ \hat{U}_{2}\right\} =\mathbb{E}\left\{ \hat{U}_{1}\right\} $.
For the expected value of $\hat{U}_{3}$, we can write {\small{}
\begin{align*}
\mathbb{E}\left\{ \hat{U}_{3}\right\} = & \mathbb{E}_{(B,D)}\Biggl\{-2BD\times\\
 & \mathbb{E}_{(-\hat{U}_{3}/2BD)|(B,D)}\left\{ \frac{\left(A+B\right)\left(C+D\right)}{\left(A+B\right)^{2}+\left(C+D\right)^{2}}|(B,D)\right\} \Biggr\}.
\end{align*}
}Using (\ref{eq:joint MGF of Q1-Q2}), the joint \ac{MGF} of $Q_{1}=\left(A+B\right)\left(C+D\right)$
and $Q_{2}=\left(A+B\right)^{2}+\left(C+D\right)^{2}$ for given $B$
and $D$ can be expressed as (\ref{eq:joint moment of Q1Q2}) shown
on the next page.
\begin{figure*}[tbh]

\begin{align}
\mathcal{M}_{(Q_{1},Q_{2})|(B,D)}\left(s_{1},s_{2}\right) & =\frac{2}{\left(4\left(1-s_{2}\right)^{2}-s_{1}^{2}\right)^{\frac{1}{2}}}\exp\Biggl[-B^{2}-D^{2}+\frac{4B^{2}\left(1-s_{2}\right)+4BDs_{1}+4D^{2}\left(1-s_{2}\right)}{4\left(1-s_{2}\right)^{2}-s_{1}^{2}}\Biggr].\label{eq:joint moment of Q1Q2}\\
\hline \nonumber 
\end{align}
\end{figure*}
 Then, considering (\ref{eq:moments of ratio}), $\mathbb{E}\left\{ \hat{U}_{3}\right\} $
is given by (\ref{eq:E(Uhat3)}) shown on the next page.
\begin{figure*}[tbh]

\begin{align}
\mathbb{E}\left\{ \hat{U}_{3}\right\} = & \frac{-2}{\Gamma^{2}\left(\frac{1}{2}\right)}\int_{-\infty}^{\infty}\int_{-\infty}^{\infty}BD\left(\int_{0}^{\infty}\frac{BD}{\left(1+s_{2}\right)^{3}}\exp\left(-B^{2}-D^{2}+\frac{B^{2}}{1+s_{2}}+\frac{D^{2}}{1+s_{2}}\right)\mathrm{d}s_{2}\right)\nonumber \\
 & \exp\left(-B^{2}\right)\exp\left(-D^{2}\right)\mathrm{d}B\mathrm{d}D\nonumber \\
= & \frac{-2}{\Gamma^{2}\left(\frac{1}{2}\right)}\int_{0}^{\infty}\frac{1}{\left(1+s_{2}\right)^{3}}\left(\int_{-\infty}^{\infty}B^{2}\exp\left(-B^{2}\frac{1+2s_{2}}{1+s_{2}}\right)\mathrm{d}B\right)\left(\int_{-\infty}^{\infty}D^{2}\exp\left(-D^{2}\frac{1+2s_{2}}{1+s_{2}}\right)\mathrm{d}D\right)\mathrm{d}s_{2}\nonumber \\
= & -\frac{1}{2}\int_{0}^{\infty}\frac{1}{\left(1+2s_{2}\right)^{3}}\mathrm{d}s_{2}=-\frac{1}{8}.\label{eq:E(Uhat3)}\\
\hline \nonumber 
\end{align}
\end{figure*}

Therefore, $\mathbb{V}\left\{ \hat{Y}_{i}\right\} $ is derived as
\[
\mathbb{V}\left\{ \hat{Y}_{i}\right\} =\mathbb{E}\left\{ \hat{Y}_{i}^{2}\right\} =2\cdot\frac{3}{16}-\frac{1}{8}=\frac{1}{4}.
\]

This is then used to determine the variance of $Z_{2}$, thus completing
the proof of Theorem \ref{thm:case-(ii)}.

\section{Proof of Theorem \ref{thm:case-(iii)}\label{sec:proof of theorem case(iii)}}

Here we analyze the expected value and variance of $Y_{i}$ and $\hat{Y}_{i}$
where $m=n$ and $\hat{m}\neq n$ (case (iii)).

In this case, the expected value and variance of $\hat{Y}_{i}$ are
the same as the corresponding values derived in case (i). Since $m=n$,
we have $A_{i}=D_{i}$ and $C_{i}=-B_{i}$. Hence, $Y_{i}$ is given
by 
\[
Y_{i}=\frac{1}{\sqrt{2}}\sqrt{A_{i}^{2}+B_{i}^{2}},
\]
which is a Rayleigh \ac{RV} with $\mathbb{E}\left\{ Y_{i}\right\} =\frac{\sqrt{\pi}}{2\sqrt{2}}$
and $\mathbb{V}\left\{ Y_{i}\right\} =\frac{4-\pi}{8}$. The proof
of Theorem \ref{thm:case-(iii)} then follows.

\bibliographystyle{ieeetr}
\bibliography{ref}

\end{document}

%% file: acro.tex
\acrodef{5G}{fifth-generation}
\acrodef{6G}{sixth-generation}
\acrodef{ABEP}{average bit error probability}
\acrodef{AO}{alternating optimization}
\acrodef{APGM}{accelerated projected gradient method}
\acrodef{ASP}{antenna separation product}
\acrodef{AWGN}{additive white Gaussian noise}
\acrodef{BEP}{bit error probability}
\acrodef{BER}{bit error rate}
\acrodef{BF-MIMO}[BF\mbox{-}MIMO]{beamforming MIMO}
\acrodef{BF}{beamforming}
\acrodef{bpcu}{bits per channel use}
\acrodef{CDF}{cumulative distribution function}
\acrodef{CF}{characteristic function}
\acrodef{CLT}{central limit theorem}
\acrodef{CP}{cyclic prefix}
\acrodef{CSI}{channel state information}
\acrodef{CSIR}{channel state information at receiver}
\acrodef{CSIT}{channel state information at the transmitter}
\acrodef{DCMC}{discrete\mbox{-}input continuous\mbox{-}output memoryless channel}
\acrodef{DFT}{discrete Fourier transform}
\acrodef{DL-TR-GSM}{dual-layered transmit-receive \ac{GSM}}
\acrodef{DLT}{dual-layered transmission}
\acrodef{EGC}{equal gain combining}
\acrodef{EM}{electromagnetic}
\acrodef{FSPL}{free space path loss}
\acrodef{FFT}{fast Fourier transform}
\acrodef{FDE}{frequency domain equalization}
\acrodef{GD}{greedy detector}
\acrodef{GRSM}{generalized \acl{RSM}}
\acrodef{GSM}{generalized \acl{SM}}
\acrodef{IFFT}{invserse fast Fourier transform}
\acrodef{ICI}{inter-channel interference}
\acrodef{iid}[i.i.d.]{independent and identically distributed}
\acrodef{IM}{index modulation}
\acrodef{IQ}{in\mbox{-}phase and quadrature}
\acrodef{ISI}{intersymbol interference}
\acrodef{ISI-free}[ISI\mbox{-}free]{intersymbol interference free}
\acrodef{KKT}{Karush\mbox{-}Kuhn\mbox{-}Tucker}
\acrodef{LDS}{low density signature}
\acrodef{LIS}{large intelligent surface}
\acrodef{LOS}{line\mbox{-}of\mbox{-}sight}
\acrodef{LT}{Laplace transform}
\acrodef{mmWave}{millimeter-wave}
\acrodef{mMIMO}{massive \ac{MIMO}}
\acrodef{MGF}{moment generating function}
\acrodef{MIMO}{multiple\mbox{-}input multiple\mbox{-}output}
\acrodef{MISO}{multiple\mbox{-}input single\mbox{-}output}
\acrodef{ML}{maximum likelihood}
\acrodef{MP}{message\mbox{-}passing}
\acrodef{MRC}{maximal ratio combining}
\acrodef{MMSE}{minimum mean square error}
\acrodef{MSTBC}{multi\mbox{-}dimensional \ac{STBC}}
\acrodef{MU-TR-GSM}{multiuser transmit-receive  \ac{GSM} }
\acrodef{NCSIT}{no channel state information at TX}
\acrodef{NLOS}{non\mbox{-}\acs{LOS}} 
\acrodef{NOMA}{non-orthogonal multiple access}
\acrodef{OFDM}{orthogonal frequency division multiplexing}
\acrodef{OFDMA}{orthogonal frequency division multiple access}
\acrodef{PA}{power amplifier}
\acrodef{PAE}{power added efficiency}
\acrodef{PAPR}{peak\mbox{-}to\mbox{-}average power ratio}
\acrodef{PDF}{probability density function}
\acrodef{PEP}{pairwise error probability}
\acrodefplural{PEP}{pairwise error probabilities}
\acrodef{PGM}{projected gradient method}
\acrodef{PMP}{probability mass function}
\acrodef{PSM}{precoding-aided spatial modulation}
\acrodef{QSM}{quadrature spatial modulation}
\acrodef{QSSK}{quadrature space-shift keying}
\acrodef{RC}{reorganization computation}
\acrodef{RE}{resource element}
\acrodef{RF}{radio frequency}
\acrodef{RIS}{reconfigurable intelligent surface}
\acrodef{RIS-AP}{RIS-access point}
\acrodef{RIS-RQRM}{RIS\mbox{-}aided receive quadrature reflecting modulation}
\acrodef{RIS-RQSSK}[\ac{RIS}\mbox{-}RQSSK]{\ac{RIS}\mbox{-}assisted receive quadrature space-shift keying}
\acrodef{RIS-SM}{\ac{RIS}-\acl{SM}}
\acrodef{RIS-SSK}{\ac{RIS}\mbox{-}\acl{SSK}}
\acrodef{RPM}{reflection pattern modulation}
\acrodef{RSM}{receive spatial modulation}
\acrodef{RV}{random variable}
\acrodef{RX}{receiver}
\acrodef{SCMA}{sparse code multiple access}
\acrodef{SEP}{symbol error probability}
\acrodef{SER}{symbol error rate}
\acrodef{SINR}{signal-to-interference-plus-noise ratio}
\acrodef{SISO}{single-input single-output}
\acrodef{SL}{sparse layering}
\acrodef{SL-MIMO}{sparse layered \ac{MIMO}}
\acrodef{SM}{spatial modulation}
\acrodef{SMX-MIMO}[SMX\mbox{-}MIMO]{spatial multiplexing MIMO}
\acrodef{SMX}{spatial multiplexing}
\acrodef{SNR}{signal-to-noise ratio}
\acrodef{SC}{single carrier}
\acrodef{SVD}{singular value decomposition}
\acrodef{SPST}{single pole single-throw}
\acrodef{SSK}{space-shift keying}
\acrodef{STBC}{space\mbox{-}time block code}
\acrodef{SU}{secondary user}
\acrodef{TDE}{time domain equalization}
\acrodef{TX}{transmitter}
\acrodef{ULA}{uniform linear array}
\acrodef{URA}{uniform rectangular array}
\acrodef{VBLAST}{vertical Bell Labs layered space\mbox{-}time}
\acrodef{VGA}{variable gain amplifier}
\acrodef{ZF}{zero-forcing}
\acrodef{ZMCG}{zero-mean complex Gaussian}